\documentclass{nature}

\usepackage{graphicx}

\makeatletter
\let\saved@includegraphics\includegraphics
\AtBeginDocument{\let\includegraphics\saved@includegraphics}
\renewenvironment*{figure}{\@float{figure}}{\end@float}
\renewenvironment*{table}{\@float{table}}{\end@float}
\makeatother

\usepackage{siunitx}
\usepackage{xstring}
\usepackage{etoolbox}
\usepackage{caption}

\makeatletter
\newcommand\formatlabel[1]{%
    \noexpandarg
    \IfSubStr{#1}{.}{%
      \StrBefore{#1}{.}[\firstcaption]%
      \StrBehind{#1}{.}[\secondcaption]%
      \textbf{\firstcaption.} \secondcaption}{%
      #1}%
      }

\patchcmd{\@caption}{#3}{\formatlabel{#3}}
\makeatother

\title{Tension tuning of sound and heat transport in graphene}

\author{H. Liu$^{1}$, M. Lee$^{2}$, M. \v{S}i\v{s}kins$^{1,2}$, H. S. J. van der Zant$^{2}$, P. G. Steeneken$^{1,2}$ and G. J. Verbiest$^1$}

\begin{document}

\maketitle

\begin{affiliations}
 \item Department of Precision and Microsystems Engineering, Delft University of Technology
 \item Kavli Institute of Nanoscience, Delft University of Technology
\end{affiliations}

\begin{abstract}%

Heat transport by acoustic phonons in 2D materials is fundamentally different from that in 3D crystals\cite{chen2012thermal,lindsay2010flexural,wang2020distinguishing,lee2015hydrodynamic} because the out-of-plane phonons propagate in a unique way that strongly depends on tension and bending rigidity. Since in-plane and out-of-plane phonon baths are decoupled, initial studies suggested they provide independent pathways for heat transport and storage in 2D materials\cite{sullivan2017optical,dolleman2020nonequilibrium}. Here, we induce tension in freestanding graphene membranes by electrostatic force, and use optomechanical techniques to demonstrate that it can change the rate of heat transport by as much as 33$\%$. Using a ballistic Debye model, we account for these observations and extract the average bending rigidity of the flexural acoustic phonons, which increases approximately linearly with the membrane’s areal mass density, in contrast to the cubic dependence seen in bulk structures. Thus, we not only elucidate phononic heat transport mechanisms in suspended 2D materials, but also provide a promising route for controlling nanoscale heat transport by tension.

\end{abstract}


Although in most bulk materials the propagation speed of different types of acoustic phonons is of similar magnitude, the situation is vastly different in 2D materials\cite{taheri2021importance,cepellotti2015phonon}. In these atomically thin materials, in-plane phonons have a propagation speed that is determined by the atomic bond stiffnesses, whereas out-of-plane flexural phonons, exhibit a speed that is dominated by tension and can be more than an order of magnitude smaller\cite{mariani2008flexural,castro2010limits,bonini2012acoustic,tornatzky2019phonon}, while also showing a unique quadratic dispersion relation.    


Evidence for the importance of these speed differences on the phononic heat transport in 2D materials was provided by theoretical analysis\cite{vallabhaneni2016reliability} and by the observation of two distinct thermal time constants in graphene membranes, of which the longest, $\tau$, is a probe for studying heat transport by the relatively slow flexural phonons\cite{dolleman2020nonequilibrium,dolleman2018transient}. To understand heat propagation via these different in-plane and out-of-plane phononic channels in 2D materials, studies of the role of flexural phonons are essential. Yet, unlike the thermal conductivity of 2D materials which has been well characterized by Raman microscopy\cite{ghosh2008extremely,balandin2008superior}, a microscopic picture of how the rate of heat transport is related to the properties of flexural phonons remains elusive, as it requires a methodology for measuring their effect on temperature variations in suspended 2D materials with nanosecond resolution. 

In this letter, we experimentally determine the heat transport rate in graphene drum resonators and demonstrate its tunability by tension. We first determine the tension and effective mass of four graphene resonators using high-frequency optomechanical technique\cite{dolleman2017optomechanics} by measuring the dependence of the mechanical resonance frequency on gate voltage induced electrostatic forces. Then, by applying the same technique in low frequency domain, we determine the gate tunability of the thermal time constant and effective thermal expansion coefficients of the membranes. Finally, by relating these measurements to a model for phononic heat transport, we characterize the ballistic transport and boundary scattering of phonons, the effect of graphene's bending rigidity on 2D heat transport and the relative thermal expansion contributions of flexural and in-plane phonons.

\begin{figure}
\centering
	\includegraphics[width=1\linewidth,angle=0]{"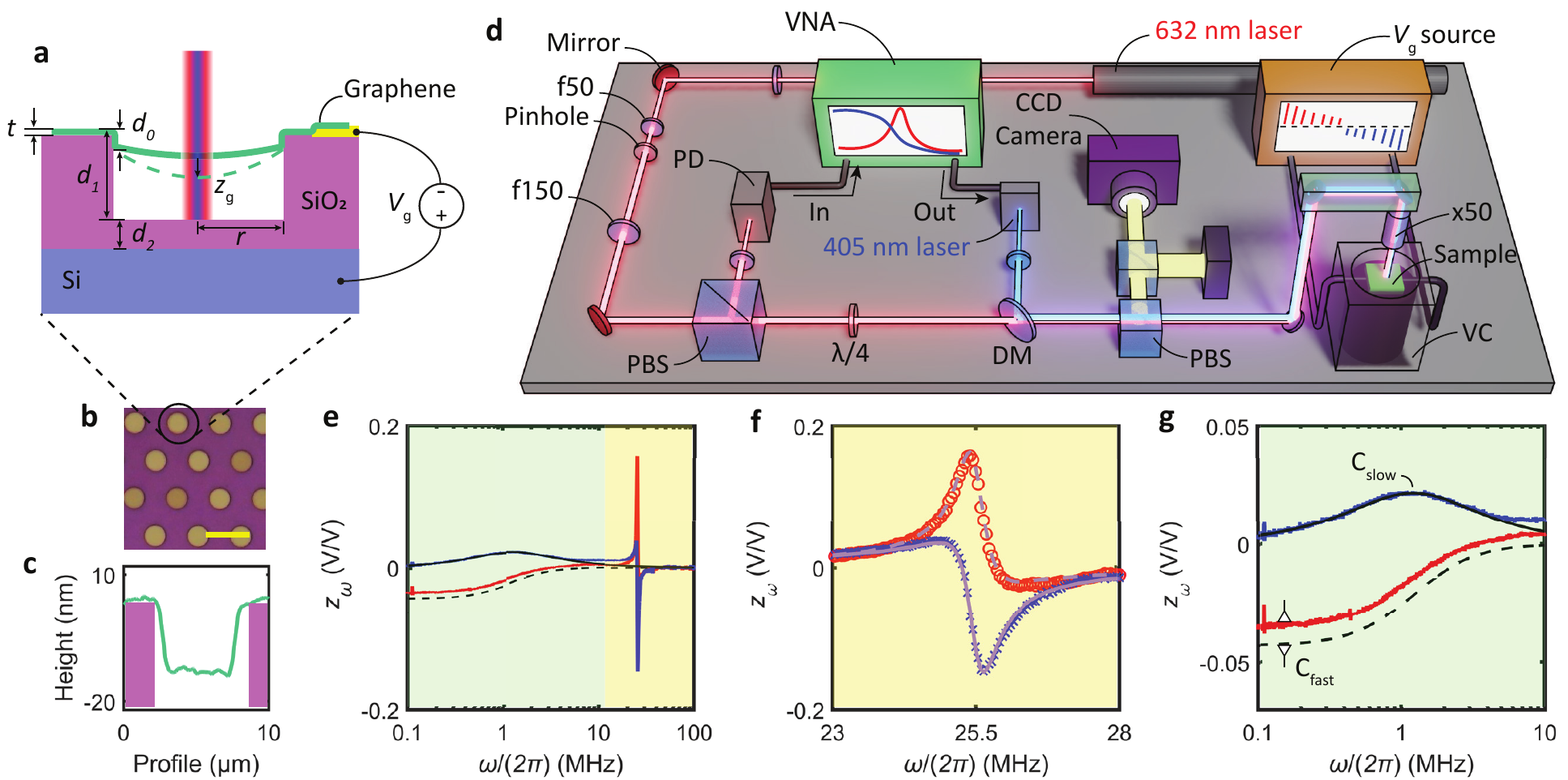"}
	\caption{\textbf{Graphene membrane characterization}. \textbf{a}, Schematic of a graphene drum of radius $r$ over a cavity of depth $d_1$ in a SiO$_2$/Si substrate irradiated by lasers. The graphene has a sidewall of length $d_0$ into the cavity. The effective electrostatic gap is expressed as $g_0=d_1-d_0+d_2/\varepsilon_{\text{{SiO}}_2}$. The voltage $V_g$ pulls the drum down by $z_g$, increasing its tension. \textbf{b}, Optical image of the drums with a scale bar of \SI{10}{\micro\meter}. \textbf{c}, Atomic force microscope line trace over a suspended drum indicates a downward deformation of drum. \textbf{d}, Interferometric setup. The sample is placed inside a vacuum chamber (VC). The blue (\SI{405}{\nano\meter}) laser is intensity modulated by a vector network analyzer (VNA) to actuate the resonator. Intensity variations of the reflected red (\SI{633}{\nano\meter}) laser caused by resonator motion, are measured by photodiode (PD) and recorded with the VNA. PBS: polarized beam splitter; DM: dichroic mirror. \textbf{e}, Frequency response of device D1, including real (red) and imaginary (blue) parts of the motion $z_\omega$. \textbf{f}, Fits of $z_\omega$ (lines) to equation (1) to obtain the resonance frequency $\omega_0$. \textbf{g}, Fits of equation (1) to $z_\omega$ near the thermal peak (black solid and dashed lines) provide the thermal time constant $\tau$ and thermal expansion $C_{\text{fast}}$ and $C_{\text{slow}}$ of the resonator.} 
	\label{fig:1}
\end{figure}

We present measurements on four CVD-grown double-layer graphene drum resonators D1$-$D4 (see fabrication in Methods). Resonators D1 and D2 have radii of $r=$ \SI{2.5}{\micro\meter} and D3 and D4 have radii of $r=$ \SI{5}{\micro\meter}. Fig.~\ref{fig:1}a-b shows a schematic illustration and an optical microscope image of a drum resonator. The cavity depth $d_1$ is \SI{245}{\nano\meter} and the thickness of the remaining unetched SiO$_2$ layer at the bottom of the cavity $d_2$ is \SI{40}{\nano\meter}. The surface profile measured by atomic force microscopy (AFM) indicates an initial downward deflection $d_0$ of the membrane, resulting from sidewall adhesion at the edge of the membrane\cite{bunch2008impermeable} (see Fig.~\ref{fig:1}c). We measure the motion of the membrane using the interferometer depicted in Fig.~\ref{fig:1}d (see Methods and SI section 1 for details).

Figure.~\ref{fig:1}e shows the measured motion amplitude $z_{\omega}$ of device D1 at $V_{\text{g}}=$ \SI{0}{\volt} over the frequency range from \SI{0.1}{} to \SI{100}{\MHz}. We extract the resonance frequency $\omega_0$ and quality factor $Q$ by fitting the measured data to a harmonic oscillator model (Fig.~\ref{fig:1}f). For device D1 this results in ${\omega_0}/{(2\pi)}=$\SI{25.49}{\mega\hertz} and $Q=43.25$. Around \SI{1}{\MHz}, we observe an additional broad peak in the imaginary part of the frequency response (Fig.~\ref{fig:1}g). We identify this peak as the thermal peak of resonator, since it is only there when driving the membrane optothermally\cite{dolleman2017optomechanics}. Following literature\cite{dolleman2020nonequilibrium}, below mechanical resonance, the displacement $z_{\omega}$ is a measure of the displacement due to thermal expansion that is delayed with respect to the laser power as a consequence of the time it takes to increase the membrane temperature by laser heating. Around the thermal peak in Fig.~\ref{fig:1}g, $z_{\omega}$ is given by:
\begin{equation}
\label{eq 1}
\ z_{\omega}=\frac{C_{\text{slow}}}{1+i{\omega}{\tau}}+C_{\text{fast}}, 
\end{equation}
where $\tau$ is the thermal time constant, and $C_{\text{slow}}$ and $C_{\text{fast}}$ quantify the device's effective thermal expansion coefficients due to the out-of-plane and in-plane phonons, respectively\cite{dolleman2020nonequilibrium}. We extract these parameters by fitting the real and imaginary parts of the measured $z_{\omega}$ with equation~\ref{eq 1}, as depicted in Fig.~\ref{fig:1}g. Here, the thermal peak is located at \SI{1.23}{\mega\hertz}, corresponding to $\tau =(2 \pi \times 1.23\mathrm{~MHz})^{-1}=$ \SI{129}{\nano\second}. 

\begin{figure}
\centering
	\includegraphics[width=0.9\linewidth,angle=0]{"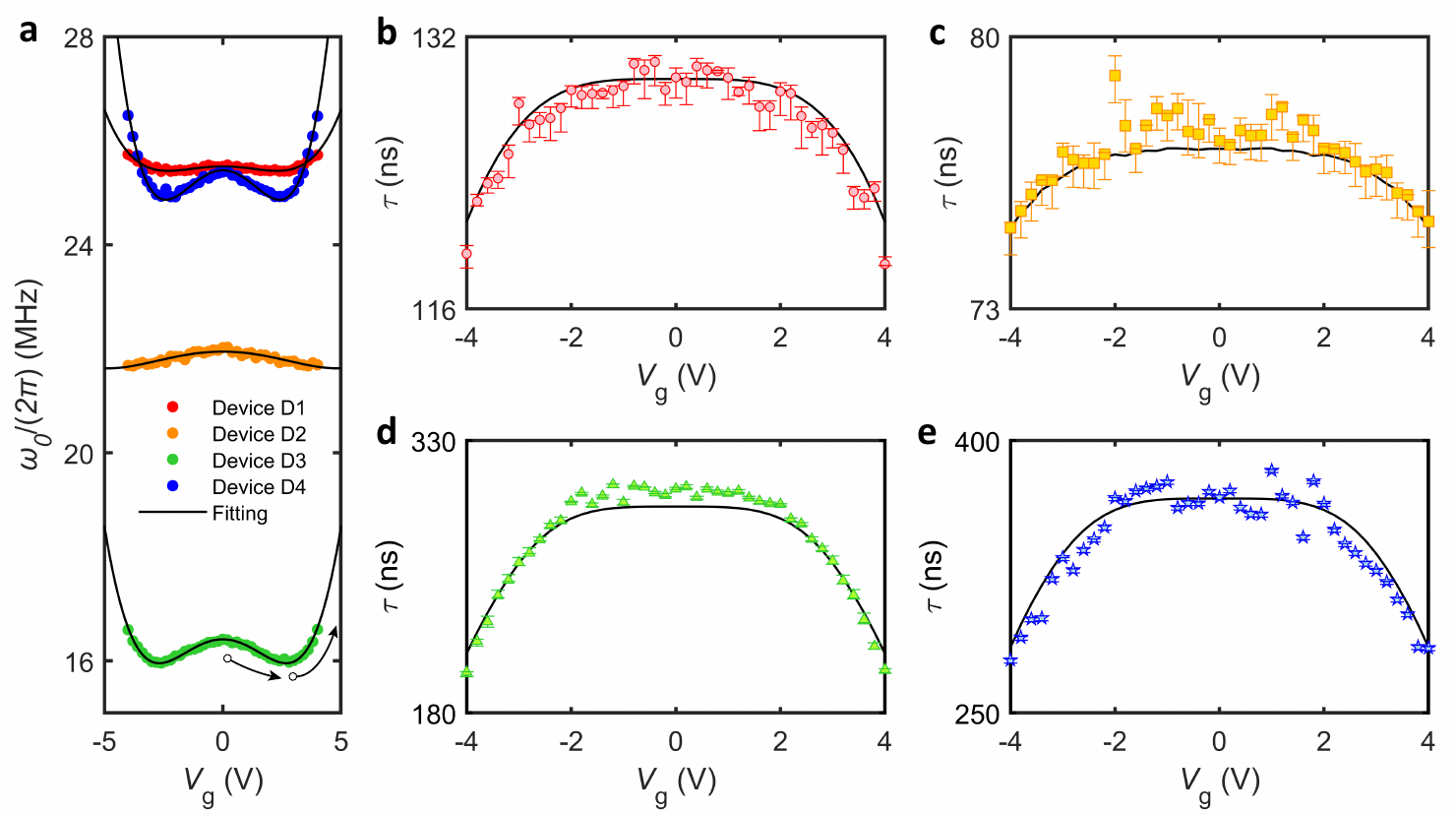"}
    \caption{\textbf{Tuning thermodynamic properties of graphene drum devices with gate voltage}. \textbf{a}, Solid dots: resonance frequency ${\omega_0}$ versus gate voltage $V_{\text{g}}$ measured in devices D1$-$D4; drawn lines: fits based on equation (2); arrows indicate the modulation from the capacitive softening regime to the tension dominated (hardening) regime. \textbf{b}$-$\textbf{e}, Points: thermal time constant $\tau$ versus $V_{\text{g}}$ measured in devices D1$-$D4, respectively; solid lines: fits to data using the Debye-scattering model; error bars are from the fits to the measured thermal peaks as plotted in Fig.~\ref{fig:1}g.}
	\label{fig:2}
\end{figure}

To determine the tension $n$ in the drum, we measure its resonance frequency ${\omega_0}$ as a function of applied electrostatic gate voltage $V_{\text{g}}$ (see Fig.~\ref{fig:1}a). The voltage $V_{\text{g}}$ generates an electrostatic force, pulling the drum down thereby inducing tension. Figure.~\ref{fig:2}a shows ${\omega_0}/{(2\pi)}$ against $V_{\text{g}}$ for all devices from \SI{-4}{\volt} to \SI{4}{\volt}. The typical W-shaped curves show both electrostatic softening and tension hardening (indicated by arrows), as often observed in electrostatic gate-tuning of graphene NEMS\cite{chen2009performance}. Following literature\cite{chen2009performance}, we model the frequency tuning of the drum resonator from continuum mechanics with:
\begin{equation}
\label{eq 3}
\ \omega_0(V_{\text{g}}) = \sqrt{\frac{1}{m_{\text{eff}}} 	\bigg[ \frac{2\pi Et \epsilon_0}{1-\upsilon^2}+\frac{8\pi Et}{(1-\upsilon^2)r^2}{z_{\text{g}}^2}-\frac{1}{2}\frac{\partial^2C_{\text{g}}}{\partial z_\text{g}^2}V^2_{\text{g}} \bigg] },\ 
\end{equation}
in which $\epsilon_0$ is the built-in strain, $m_{\text{eff}}$ is the modal mass of the fundamental mode of the circular membrane resonator with a theoretical value $m_{\text{eff,th}}=0.271\pi r^2 \rho_g$ where $\rho_g$ is the mass density of double-layer graphene, $C_{\text{g}}$ is the capacitance between membrane and bottom gate, and the derivative $\frac{\partial^2C_{\text{g}}}{\partial z_{\text{g}}^2}$ quantifies the electrostatic softening. The 2D Young's modulus $Et \approx$\SI{175.39}{\newton/\meter} and Poisson ratio $\upsilon=0.16$ were determined via AFM indentation\cite{vsivskins2020sensitive} (see SI section 1). The center deflection $z_{\text{g}}$ can be expressed\cite{weber2014coupling} as ${\varepsilon_0 r^2V_{\text{g}}^2}/{(8g_0^2 n_0)}$, where $n_0={Et\epsilon_0}/({1-\upsilon})$ is the pretension, $\varepsilon_0$ is the permittivity of vacuum, and $g_0$ is the effective gap between the drum and the electrostatic gate. We fit the measured $V_{\text{g}}$ dependence of ${\omega_0}$ with equation (2) (black lines in Fig. 2a) to extract $n_0$, $m_{\text{eff}}$ and $\frac{\partial^2C_{\text{g}}}{\partial z^2}$ for each device (listed in Table 1).

The extracted initial tension $n_0$ ranging between \SI{0.18}{} to \SI{0.34}{N/m} is in agreement with the literature values reported for graphene NEMS\cite{chen2009performance,weber2014coupling,will2017high,zhang2020dynamically}, and $m_{\text{eff}}$ of \SI{0.39}{} to \SI{1.06}{} $\times10^{-16}$ \SI{}{kg} is larger than the mass expected for double-layer graphene, which is attributed to polymer residues left after fabrication\cite{miao2014graphene,will2017high}. The values for $\frac{\partial^2C_{\text{g}}}{\partial z_{\text{g}}^2}$ are close to the theoretical estimate $0.542{\varepsilon_0 \pi r^2}/{g_0^3}$ from the geometry of the capacitor\cite{vsivskins2020magnetic}. The excellent agreement between the fit and measured dependence of ${\omega_0}$ on gate voltage (Fig.~\ref{fig:2}a) allows to extract the membrane tension $n$ at each value of $V_g$ and corresponding center deflection $z_g$, using the equation\cite{chen2009performance} $n=n_0(1+{4z_\text{g}^2}/{r^2})$.

Besides the changes in the resonance frequency ${\omega_0}$, we also observe that the measured thermal time constant $\tau$ decreases with $V_{\text{g}}$ for all devices (dots, Fig. 2b$-$2d), as much as 33$\%$. To shed light on this observation, we use a phononic model to assess the transient thermal conduction in the membrane\cite{vallabhaneni2016reliability}. We assume in the model that the transport of flexural acoustic phonons in the membrane is ballistic, since their mean free paths are much longer than the radius $r$ of membranes\cite{fugallo2014thermal}, while using a scattering model\cite{dolleman2020phonon} (see Methods and Fig. S1) to capture the tension dependent phonon transport across the kink at the edge of the membrane (Fig. \ref{fig:1}c). This results in the following function: 
\begin{equation}
\label{eq 2}
\ \tau=\frac{r}{2\sum_j\overline{w}_{1z\to{2j}}c_z}, j=l,t,z,  
\end{equation}
in which $\overline{w}_{1z\to{2j}}$ is the probability that a flexural phonon on the suspended part of the graphene (subscript 1) is transmitted across the membrane edge and becomes a phonon of type $j$ on the supported part of the graphene (subscript 2), where $j$ can either be a flexural phonon ($j=z$), or a longitudinal or transverse in-plane phonon ($j=l,t$). The scattering probability $\overline{w}_{1z\to{2j}}$ depends both on tension $n$ and on the speed of sound $c_z$ of flexural phonons. The dispersion relation for flexural phonons is given by $\omega_q = \sqrt{(\kappa q^4+n q^2)/(\eta\rho_g)}$ (ref.\cite{nihira2003temperature}),  where $q$ is the wavenumber, $\kappa$ is the bending rigidity of the membrane, and $\eta = m_{\text{eff}}/(0.271\pi r^2 \rho_g)$ is the normalized areal mass of the membrane (see Table 1). From the dispersion relation the speed of sound for flexural phonons can be found using $c_z=\frac{\partial \omega_q}{\partial q}$.  

\begin{figure}
\centering
	\includegraphics[width=0.75\linewidth,angle=0]{"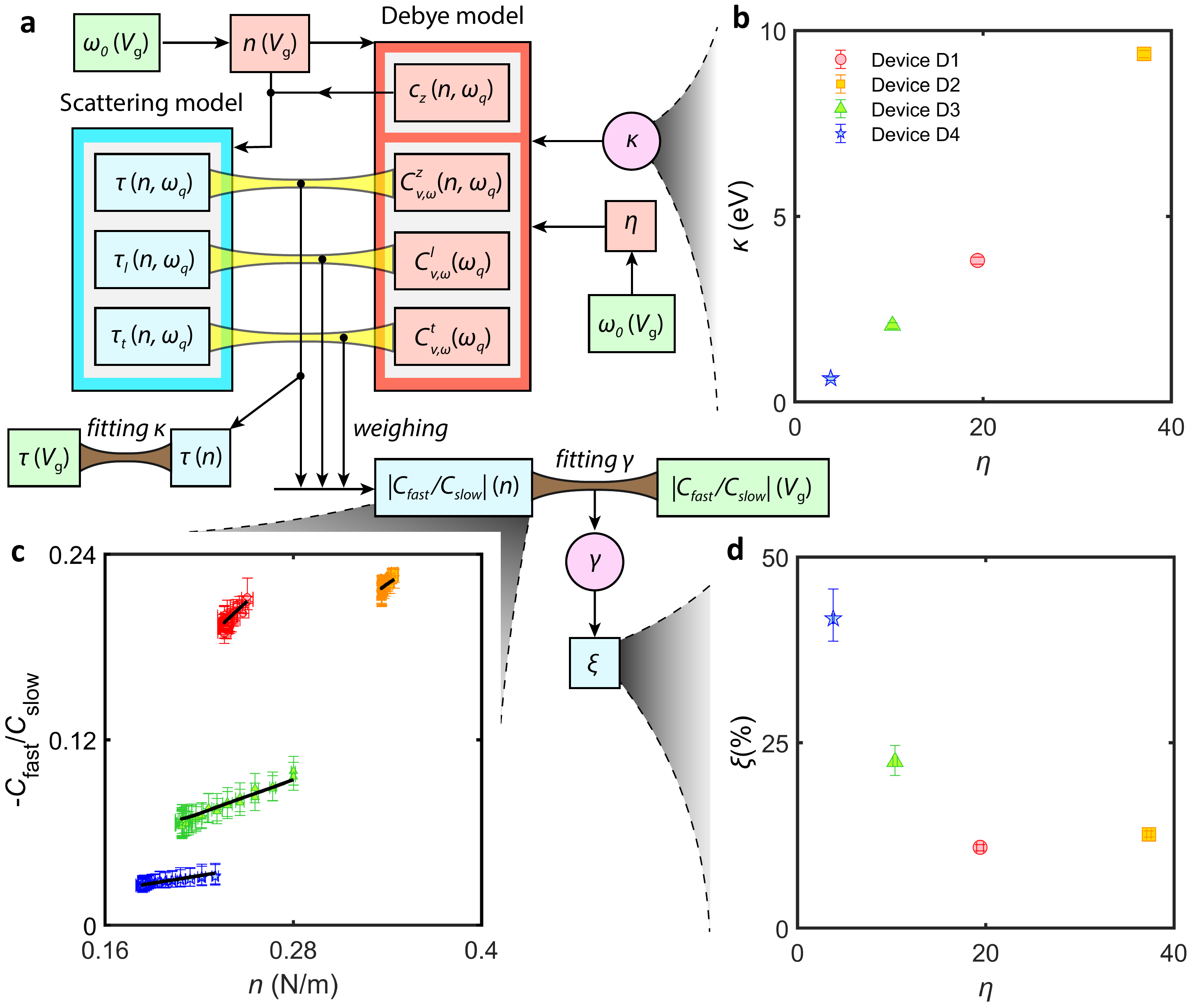"}
    \caption{\textbf{Analysis and experimental demonstration on tunable heat transport in graphene drum devices}. \textbf{a}, Flow chart of the theoretical model to estimate the thermal time constant $\tau$ and thermal expansion ratio $|C_{\text{fast}}/C_{\text{slow}}|$. Green frames: measured parameters; pink frames: fitting parameters. From Debye model we obtain the phonon frequency $\omega_q$- dependent speed $c_z$ and specific heats $C_v^z$, $C_v^l$, $C_v^t$ for phonons, which are used as weighing factors in the scattering model to evaluate $\tau$ and $|C_{\text{fast}}/C_{\text{slow}}|$ as a function of tension $n$. \textbf{b}, Fitted bending rigidity $\kappa$ of the membrane versus normalized areal mass $\eta$. These values of $\kappa$ result in a good match between the modelled and measured $\tau$ for devices D1$-$D4, as plotted in Fig.~\ref{fig:2}b$-$2e (drawn lines). \textbf{c}, $|C_{\text{fast}}/C_{\text{slow}}|$ versus $n$. Points: the measurements; solid lines: the modelled estimates; error bars are from the fits to the thermal peaks. \textbf{d}, Percentual energy absorption $\xi$ by flexural phonons as a function of $\eta$ shows a strong decrease of $\xi$ with increasing mass due to polymer residues. }
	\label{fig:3}
\end{figure}

In practice, flexural phonons with frequencies up to the Debye frequency $\omega_{qd}$, will contribute to the heat transport and thermal time constant $\tau$. To account for this (see Fig.~\ref{fig:3}a), we integrate over the Bose-Einstein distribution using the phonon speed distribution $c_z(\omega_q,n)$ to calculate the specific heat spectral density $C_{v,\omega}^{z}({\omega_q},n)$ of flexural phonons (Debye model); then, knowledge of the tension $n$ and speed of sound $c_{z}$ allows to solve equation (3) and determine $\tau({\omega_q},n)$; finally a weighted average over the contribution of phonons at each frequency $\omega_q$ to the thermal time constant $\tau(n)$ is determined using ${1}/{\tau(n)}= \int_{0}^{\omega_{qd}} C_{v,\omega}^{z}({\omega_q},n)/({C_v^{z}(n) \tau(\omega_q,n)}) \text{d}\omega_q$, where the total specific heat due to flexural phonons is determined using $C_v^z(n) =\int_{0}^{\omega_{qd}} C_{v,\omega}^{z}({\omega_q},n) \text{d}\omega_q$\cite{nihira2003temperature,singh2011mechanism}. More details about the scattering and Debye models can be found in SI section 3 and 4.  




From the presented measurements, all parameters in the above model (green color in Fig. \ref{fig:3}a) can be determined. Only the bending rigidity of graphene is not measured, which especially affects high frequency ($\sim$ THz) flexural phonons near the Debye frequency $\omega_{qd}$ (Fig. S4). We therefore use the bending rigidity $\kappa$ as a fit parameter, to match the modelled $\tau$ to the experimental values for all devices D1$-$D4 (see Fig.~\ref{fig:2}b$-$2e). The value of $\kappa$ varies in a broad range, from 0.6 to \SI{9.4}{eV} roughly, which is comparable to values reported in literature\cite{han2020ultrasoft,wei2013bending}. Interestingly, $\kappa$ is roughly proportional to the normalized areal mass $\eta$ of the membranes (see Fig.~\ref{fig:3}b), indicating that polymer residues not only affect the mechanical properties of the membrane, but also seem to play an important role in the heat transport by increasing the membrane's bending rigidity.         
 
Let us now focus on the amplitudes $C_{\text{fast}}$ and $C_{\text{slow}}$ in equation (1), which measure the magnitude of the thermal expansion of the in-plane and flexural phonons, respectively (as indicated in Fig.~\ref{fig:1}g). The measured ratio $|C_{\text{fast}}/C_{\text{slow}}|$ is approximately proportional to the relative temperature increase of both phonon baths\cite{dolleman2020nonequilibrium} and increases with tension $n$ from 0.02 to 0.23 (Fig.~\ref{fig:3}c, points). Since the temperature increase depends both on the absorbed laser power and the cooling rate by heat transport, we have $|C_{\text{fast}}/C_{\text{slow}}| = \gamma \frac{(\tau_l/C_v^l+\tau_t/C_v^t)}{\tau/C_v^z}$, where $\gamma = P_{\text{abs}}^{{l,t}}/P_{\text{abs}}^{{z}}$ is the power absorption ratio between in-plane and flexural phonons (see SI section 5 for derivation). By applying the same procedure as for $\tau$, we calculate $\tau_l$ and $\tau_t$ for the in-plane phonons using their specific heats $C_{v}^{l}$ and $C_{v}^{t}$ (see Fig. S3). From these values we determine the  ratio $|{C_{\text{fast}}}/{C_{\text{slow}}}|$ from the model and use $\gamma$ to fit it to the measured thermal expansion ratio, obtaining a single value of $\gamma$ for each device (listed in Table 1). As shown in Fig.~\ref{fig:3}c the model accurately captures the tension dependence of the ratio $|{C_{\text{fast}}}/{C_{\text{slow}}}|$ for all devices. The observed increase in $|{C_{\text{fast}}}/{C_{\text{slow}}}|$ is a consequence of $c_z$ increase with increasing tension, which reduces $\tau$ such that $C_{\text{slow}}$ decreases\cite{castro2010limits}. From $\gamma$, we can calculate the percentage $\xi$ of optical power that is converted into flexural phonons, expressed as $\xi=P_{\text{abs}}^{{z}}/(P_{\text{abs}}^{{l,t}}+P_{\text{abs}}^{{z}})\times 100\% = (1+\gamma)^{-1}\times 100\%$. As Fig.~\ref{fig:3}d shows, $\xi$ rapidly decreases from 43.7\% to less than 15\% when the normalized areal mass $\eta$ increases.

\begin{figure}
\centering
	\includegraphics[width=0.75\linewidth,angle=0]{"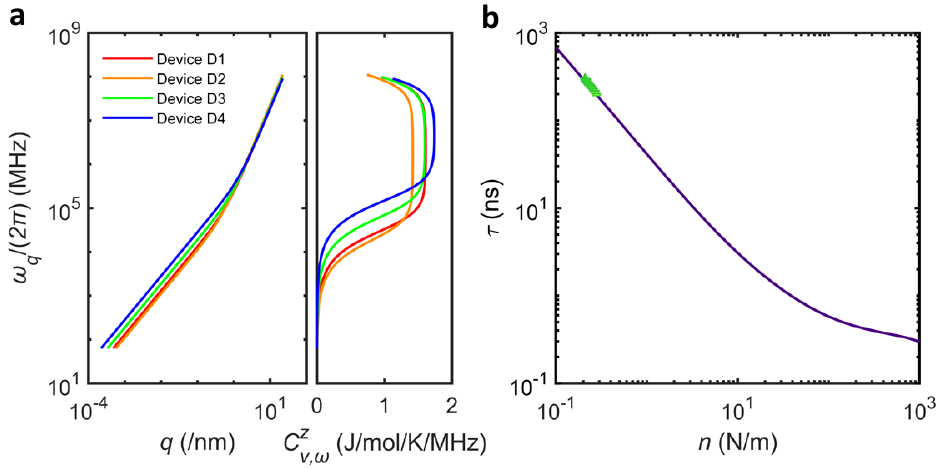"}
    \caption{ \textbf{Role of surface tension played on flexural phonons and tunability of heat transport rate in graphene membrane}. \textbf{a}, Left panel: calculated dispersion relation of out-of-plane flexural phonons $\omega_q(q)$ for devices D1$-$D4, where the bending rigidity $\kappa$ and normalized areal mass $\eta$ are used from Table 1. $\omega_q(q)$ is dominated by tension $n$ in $\sim$MHz regime and dominated by $\kappa$ in $\sim$THz regime, respectively, while the cross-over frequency $\omega_{qc}$ is located at $\sim$GHz regime. Right panel: specific heat spectral density $C_{v,\omega}^z$ for all devices. \textbf{b}, Tunable thermal time constant $\tau$ with $n$ varying from 0.1 to \SI{1000}{N/m}, using device D3 as an example. Green points: measured $\tau$ versus $n$ for device D3; solid line: estimated $\tau$ by Debye scattering model.}
	\label{fig:4}
\end{figure}

In Figs. \ref{fig:2} and \ref{fig:3}c, we show that the presented model can capture the tension dependence of the resonance frequency $\omega_0$, thermal time constant $\tau$ and thermal expansion ratio $|{C_{\text{fast}}}/{C_{\text{slow}}}|$ very well for all devices, using only $\kappa$ and $\gamma$ as fit parameter. The extracted $\kappa$ shows large variation between our devices (Fig. \ref{fig:3}b) that is correlated to the areal mass density $\eta$. Since the devices are fabricated identically, we attribute the differences to variations in residues on and between the double CVD graphene layers. These residues affect acoustic phonon transport in three different ways. First, residues decrease $\xi$, the relative optical power absorbed by flexural phonons (Fig. \ref{fig:3}d), and thus decrease their contribution to the thermal expansion of graphene\cite{mercado2021impact,pettes2011influence}. Second, the effective mass of the membrane and thus of the flexural phonons, quantified by the normalized areal mass $\eta$, is increased (Table 1). Third, the bending rigidity $\kappa$ increases (Fig. \ref{fig:3}b). This results in the observed linear relation between $\kappa$ and $\eta$ which is somewhat surprising, since for a bulk material $\kappa$ is expected to exhibit cubic scaling with thickness\cite{lindahl2012determination}. However, if the interlayer shear interaction between the layers is weak, a linear dependence is predicted to occur\cite{wang2019bending}. For high frequency ($\sim$THz) phonons, the variations in bending rigidity have little effect on the dispersion relation $\omega_q(q)$ for all devices D1$-$D4 (Fig.~\ref{fig:4}a, left panel). In contrast, the specific heat spectral density $C_{v,\omega}^{z}(\omega_q)$ still shows significant device-to-device variations (Fig.~\ref{fig:4}a, right panel), that are held responsible for the measured variations in the heat transport rate $\tau$. More details about how $\kappa$ and $\eta$ affect $\omega_q(q)$ and $C_{v,\omega}^{z}(\omega_q)$ can be found in SI section 4. For low frequency (MHz) flexural phonons, the situation is completely different. The resonance frequency of the graphene membranes can be understood as a standing wave of a flexural acoustic phonons and is thus proportional to the ratio of the speed of sound $c_z$ and the membrane radius, such that $\kappa$ does not play an important role. For MHz frequencies, $c_z$ is thus fully determined by $n$ and $\eta$, in-line with all experimental graphene resonators reported in the literature\cite{chen2009performance,weber2014coupling}. We estimate the cross-over frequency $\omega_{qc}$ where the phonon frequency $\omega_q(q)$ shows a transition from tension $n$- dominated to bending rigidity $\kappa$- dominated regime, at around 84.8, 52.6, 174.4 and 422.7~GHz for devices D1$-$D4, respectively (Fig. S4b). By measuring the resonance frequencies we obtain information on the tension and mass of the membrane, that can be used to determine the low frequency phonon dispersion, while with the thermal time constant and thermal expansion ratio we can estimate the bending rigidity that is essential for the high frequency phonon dispersion. Thus the measurements and model allow us to assess the acoustic phonon dispersion from the MHz to the THz regime.

For each of our devices, tension $n$ tunes the heat transport rate in three different ways: speed of sound $c_z$, the specific heat spectral density $C_{v,\omega}^{z}(\omega_q)$, and the phonon scattering rate as expressed in equation (3). Since $n$ only affects the dispersion relation of low frequency (MHz) phonons, its impact on the thermal time constant $\tau$ through the speed of sound $c_z$ and heat capacitance $C_{v,\omega}^{z}(\omega_q)$ is limited (Fig. S5f and i). Therefore, the observed decrease of thermal time constant $\tau$ in Fig.~\ref{fig:2} is attributed to the enhanced acoustic impedance matching between the flexural phonons on the suspended membrane and the in-plane phonons on the supported membrane, leading to faster heat dissipation. This conclusion is supported by the calculations in Fig.~\ref{fig:4}b that are based on the experimental device parameters and data (green points) obtained for device D3. It shows that as $n$ increases from 0.1 to \SI{1000}{N/m}, $\tau$ decreases from $\sim$\SI{1}{\micro\second} to less than \SI{0.3}{ns}, attributed both to the increase of the phonon transmission probability $\overline{w}_{1z\to{2j}}$ of flexural phonons and to the increased speed of sound ($c_z \approx \sqrt{n/(\eta\rho_g)}$ for $n\gg\kappa q^2$). More details about the influence of $\kappa$ on $\tau$ can be found in SI section 4. Note that the extremely high tension $n=$ \SI{1000}{N/m} is only for the discussion and cannot be physically realised in graphene membranes. For small tensions the thermal time constant is mainly limited by the phonon transmission probability and bending rigidity, whereas for high tensions ($n>$ 100 N/m) it is mainly the speed of phonons and the distance they have to travel that limit heat conduction. Such low $\tau$ is comparable to the propagation time of heat flux from the centre to the boundary of membrane\cite{dolleman2017amplitude,dolleman2017optomechanics}, given by $\tau=r^2/(5\alpha_{td})$ where $\alpha_{td}$ is the thermal diffusivity of the membrane, which also sets a limit on the tunable acoustic phonon transport by tension. Understanding this limit is important for proposed applications in the field of 2D phononics, such as switchable thermal transistors, ultra-sensitive thermal logic gates, and reconfigurable phononic memories\cite{maldovan2013sound,balandin2020phononics}.

Recent observations revealed that acoustic phonons play an important role in macroscopic quantum states in 2D materials, such as superconductivity and ferromagnetism\cite{liu20192d,chou2021acoustic}. We anticipate that the changes in the dispersion relation of flexural phonons found in this work could provide crucial insight into these quantum phase transitions, especially the transition temperature and doping effects in magic-angle twisted bilayer/trilayer graphene\cite{cao2018unconventional} and mutlilayer rhombohedral graphene\cite{shi2020electronic}.

In conclusion, we measured nanosecond-scale heat transport in suspended graphene drums and presented experimental demonstration of its tunability with in-plane tension. Using a Debye scattering model of acoustic phonons, we present a microscopic picture of heat transport in suspended graphene membranes, where bending rigidity and tension dominate the flexural dispersion relation for THz and MHz frequency phonons, respectively. The gained insight not only advances our fundamental understanding of acoustic phonons in 2D materials, but also enables pathways for controlled and optimized thermal management in 2D-based phononic, thermoelectric, electronic and quantum devices.

\bibliographystyle{naturemag}
\medskip
\bibliography{np-phonons-sound}

\begin{thebibliography}{10}
\expandafter\ifx\csname url\endcsname\relax
  \def\url#1{\texttt{#1}}\fi
\expandafter\ifx\csname urlprefix\endcsname\relax\def\urlprefix{URL }\fi
\providecommand{\bibinfo}[2]{#2}
\providecommand{\eprint}[2][]{\url{#2}}

\bibitem{chen2012thermal}
\bibinfo{author}{Chen, S.} \emph{et~al.}
\newblock \bibinfo{title}{Thermal conductivity of isotopically modified
  graphene}.
\newblock \emph{\bibinfo{journal}{Nat. Mater.}} \textbf{\bibinfo{volume}{11}},
  \bibinfo{pages}{203--207} (\bibinfo{year}{2012}).

\bibitem{lindsay2010flexural}
\bibinfo{author}{Lindsay, L.}, \bibinfo{author}{Broido, D.} \&
  \bibinfo{author}{Mingo, N.}
\newblock \bibinfo{title}{Flexural phonons and thermal transport in graphene}.
\newblock \emph{\bibinfo{journal}{Phys. Rev. B}} \textbf{\bibinfo{volume}{82}},
  \bibinfo{pages}{115427} (\bibinfo{year}{2010}).

\bibitem{wang2020distinguishing}
\bibinfo{author}{Wang, R.} \emph{et~al.}
\newblock \bibinfo{title}{Distinguishing optical and acoustic phonon
  temperatures and their energy coupling factor under photon excitation in nm
  2\uppercase{D} materials}.
\newblock \emph{\bibinfo{journal}{Adv. Sci.}} \textbf{\bibinfo{volume}{7}},
  \bibinfo{pages}{2000097} (\bibinfo{year}{2020}).

\bibitem{lee2015hydrodynamic}
\bibinfo{author}{Lee, S.}, \bibinfo{author}{Broido, D.},
  \bibinfo{author}{Esfarjani, K.} \& \bibinfo{author}{Chen, G.}
\newblock \bibinfo{title}{Hydrodynamic phonon transport in suspended graphene}.
\newblock \emph{\bibinfo{journal}{Nat. Commun.}} \textbf{\bibinfo{volume}{6}},
  \bibinfo{pages}{1--10} (\bibinfo{year}{2015}).

\bibitem{sullivan2017optical}
\bibinfo{author}{Sullivan, S.} \emph{et~al.}
\newblock \bibinfo{title}{Optical generation and detection of local
  nonequilibrium phonons in suspended graphene}.
\newblock \emph{\bibinfo{journal}{Nano lett.}} \textbf{\bibinfo{volume}{17}},
  \bibinfo{pages}{2049--2056} (\bibinfo{year}{2017}).

\bibitem{dolleman2020nonequilibrium}
\bibinfo{author}{Dolleman, R.~J.}, \bibinfo{author}{Verbiest, G.~J.},
  \bibinfo{author}{Blanter, Y.~M.}, \bibinfo{author}{van~der Zant, H.~S.} \&
  \bibinfo{author}{Steeneken, P.~G.}
\newblock \bibinfo{title}{Nonequilibrium thermodynamics of acoustic phonons in
  suspended graphene}.
\newblock \emph{\bibinfo{journal}{Phys. Rev. Res}}
  \textbf{\bibinfo{volume}{2}}, \bibinfo{pages}{012058} (\bibinfo{year}{2020}).

\bibitem{taheri2021importance}
\bibinfo{author}{Taheri, A.}, \bibinfo{author}{Pisana, S.} \&
  \bibinfo{author}{Singh, C.~V.}
\newblock \bibinfo{title}{Importance of quadratic dispersion in acoustic
  flexural phonons for thermal transport of two-dimensional materials}.
\newblock \emph{\bibinfo{journal}{Phys. Rev. B}}
  \textbf{\bibinfo{volume}{103}}, \bibinfo{pages}{235426}
  (\bibinfo{year}{2021}).

\bibitem{cepellotti2015phonon}
\bibinfo{author}{Cepellotti, A.} \emph{et~al.}
\newblock \bibinfo{title}{Phonon hydrodynamics in two-dimensional materials}.
\newblock \emph{\bibinfo{journal}{Nat. Commun.}} \textbf{\bibinfo{volume}{6}},
  \bibinfo{pages}{1--7} (\bibinfo{year}{2015}).

\bibitem{mariani2008flexural}
\bibinfo{author}{Mariani, E.} \& \bibinfo{author}{Von~Oppen, F.}
\newblock \bibinfo{title}{Flexural phonons in free-standing graphene}.
\newblock \emph{\bibinfo{journal}{Phys. Rev. Lett.}}
  \textbf{\bibinfo{volume}{100}}, \bibinfo{pages}{076801}
  (\bibinfo{year}{2008}).

\bibitem{castro2010limits}
\bibinfo{author}{Castro, E.~V.} \emph{et~al.}
\newblock \bibinfo{title}{Limits on charge carrier mobility in suspended
  graphene due to flexural phonons}.
\newblock \emph{\bibinfo{journal}{Phys. Rev. Lett.}}
  \textbf{\bibinfo{volume}{105}}, \bibinfo{pages}{266601}
  (\bibinfo{year}{2010}).

\bibitem{bonini2012acoustic}
\bibinfo{author}{Bonini, N.}, \bibinfo{author}{Garg, J.} \&
  \bibinfo{author}{Marzari, N.}
\newblock \bibinfo{title}{Acoustic phonon lifetimes and thermal transport in
  free-standing and strained graphene}.
\newblock \emph{\bibinfo{journal}{Nano lett.}} \textbf{\bibinfo{volume}{12}},
  \bibinfo{pages}{2673--2678} (\bibinfo{year}{2012}).

\bibitem{tornatzky2019phonon}
\bibinfo{author}{Tornatzky, H.}, \bibinfo{author}{Gillen, R.},
  \bibinfo{author}{Uchiyama, H.} \& \bibinfo{author}{Maultzsch, J.}
\newblock \bibinfo{title}{Phonon dispersion in \uppercase{M}o\uppercase{S}2}.
\newblock \emph{\bibinfo{journal}{Phys. Rev. B}} \textbf{\bibinfo{volume}{99}},
  \bibinfo{pages}{144309} (\bibinfo{year}{2019}).

\bibitem{vallabhaneni2016reliability}
\bibinfo{author}{Vallabhaneni, A.~K.}, \bibinfo{author}{Singh, D.},
  \bibinfo{author}{Bao, H.}, \bibinfo{author}{Murthy, J.} \&
  \bibinfo{author}{Ruan, X.}
\newblock \bibinfo{title}{Reliability of raman measurements of thermal
  conductivity of single-layer graphene due to selective electron-phonon
  coupling: A first-principles study}.
\newblock \emph{\bibinfo{journal}{Phys. Rev. B}} \textbf{\bibinfo{volume}{93}},
  \bibinfo{pages}{125432} (\bibinfo{year}{2016}).

\bibitem{dolleman2018transient}
\bibinfo{author}{Dolleman, R.~J.} \emph{et~al.}
\newblock \bibinfo{title}{Transient thermal characterization of suspended
  monolayer \uppercase{M}o\uppercase{S}2}.
\newblock \emph{\bibinfo{journal}{Phys. Rev. Mat.}}
  \textbf{\bibinfo{volume}{2}}, \bibinfo{pages}{114008} (\bibinfo{year}{2018}).

\bibitem{ghosh2008extremely}
\bibinfo{author}{Ghosh, D.} \emph{et~al.}
\newblock \bibinfo{title}{Extremely high thermal conductivity of graphene:
  Prospects for thermal management applications in nanoelectronic circuits}.
\newblock \emph{\bibinfo{journal}{Appl. Phys. Lett.}}
  \textbf{\bibinfo{volume}{92}}, \bibinfo{pages}{151911}
  (\bibinfo{year}{2008}).

\bibitem{balandin2008superior}
\bibinfo{author}{Balandin, A.~A.} \emph{et~al.}
\newblock \bibinfo{title}{Superior thermal conductivity of single-layer
  graphene}.
\newblock \emph{\bibinfo{journal}{Nano lett.}} \textbf{\bibinfo{volume}{8}},
  \bibinfo{pages}{902--907} (\bibinfo{year}{2008}).

\bibitem{dolleman2017optomechanics}
\bibinfo{author}{Dolleman, R.~J.} \emph{et~al.}
\newblock \bibinfo{title}{Optomechanics for thermal characterization of
  suspended graphene}.
\newblock \emph{\bibinfo{journal}{Phys. Rev. B}} \textbf{\bibinfo{volume}{96}},
  \bibinfo{pages}{165421} (\bibinfo{year}{2017}).

\bibitem{bunch2008impermeable}
\bibinfo{author}{Bunch, J.~S.} \emph{et~al.}
\newblock \bibinfo{title}{Impermeable atomic membranes from graphene sheets}.
\newblock \emph{\bibinfo{journal}{Nano lett.}} \textbf{\bibinfo{volume}{8}},
  \bibinfo{pages}{2458--2462} (\bibinfo{year}{2008}).

\bibitem{chen2009performance}
\bibinfo{author}{Chen, C.} \emph{et~al.}
\newblock \bibinfo{title}{Performance of monolayer graphene nanomechanical
  resonators with electrical readout}.
\newblock \emph{\bibinfo{journal}{Nat. Nanotechnol.}}
  \textbf{\bibinfo{volume}{4}}, \bibinfo{pages}{861--867}
  (\bibinfo{year}{2009}).

\bibitem{vsivskins2020sensitive}
\bibinfo{author}{{\v{S}}i{\v{s}}kins, M.} \emph{et~al.}
\newblock \bibinfo{title}{Sensitive capacitive pressure sensors based on
  graphene membrane arrays}.
\newblock \emph{\bibinfo{journal}{Microsyst. Nanoeng.}}
  \textbf{\bibinfo{volume}{6}}, \bibinfo{pages}{1--9} (\bibinfo{year}{2020}).

\bibitem{weber2014coupling}
\bibinfo{author}{Weber, P.}, \bibinfo{author}{Guttinger, J.},
  \bibinfo{author}{Tsioutsios, I.}, \bibinfo{author}{Chang, D.~E.} \&
  \bibinfo{author}{Bachtold, A.}
\newblock \bibinfo{title}{Coupling graphene mechanical resonators to
  superconducting microwave cavities}.
\newblock \emph{\bibinfo{journal}{Nano lett.}} \textbf{\bibinfo{volume}{14}},
  \bibinfo{pages}{2854--2860} (\bibinfo{year}{2014}).

\bibitem{will2017high}
\bibinfo{author}{Will, M.} \emph{et~al.}
\newblock \bibinfo{title}{High quality factor graphene-based two-dimensional
  heterostructure mechanical resonator}.
\newblock \emph{\bibinfo{journal}{Nano lett.}} \textbf{\bibinfo{volume}{17}},
  \bibinfo{pages}{5950--5955} (\bibinfo{year}{2017}).

\bibitem{zhang2020dynamically}
\bibinfo{author}{Zhang, X.} \emph{et~al.}
\newblock \bibinfo{title}{Dynamically-enhanced strain in atomically thin
  resonators}.
\newblock \emph{\bibinfo{journal}{Nat. Commun.}} \textbf{\bibinfo{volume}{11}},
  \bibinfo{pages}{1--9} (\bibinfo{year}{2020}).

\bibitem{miao2014graphene}
\bibinfo{author}{Miao, T.}, \bibinfo{author}{Yeom, S.}, \bibinfo{author}{Wang,
  P.}, \bibinfo{author}{Standley, B.} \& \bibinfo{author}{Bockrath, M.}
\newblock \bibinfo{title}{Graphene nanoelectromechanical systems as
  stochastic-frequency oscillators}.
\newblock \emph{\bibinfo{journal}{Nano lett.}} \textbf{\bibinfo{volume}{14}},
  \bibinfo{pages}{2982--2987} (\bibinfo{year}{2014}).

\bibitem{vsivskins2020magnetic}
\bibinfo{author}{{\v{S}}i{\v{s}}kins, M.} \emph{et~al.}
\newblock \bibinfo{title}{Magnetic and electronic phase transitions probed by
  nanomechanical resonators}.
\newblock \emph{\bibinfo{journal}{Nat. Commun.}} \textbf{\bibinfo{volume}{11}},
  \bibinfo{pages}{1--7} (\bibinfo{year}{2020}).

\bibitem{fugallo2014thermal}
\bibinfo{author}{Fugallo, G.} \emph{et~al.}
\newblock \bibinfo{title}{Thermal conductivity of graphene and graphite:
  collective excitations and mean free paths}.
\newblock \emph{\bibinfo{journal}{Nano lett.}} \textbf{\bibinfo{volume}{14}},
  \bibinfo{pages}{6109--6114} (\bibinfo{year}{2014}).

\bibitem{dolleman2020phonon}
\bibinfo{author}{Dolleman, R.~J.}, \bibinfo{author}{Blanter, Y.~M.},
  \bibinfo{author}{van~der Zant, H. S.~J.}, \bibinfo{author}{Steeneken, P.~G.}
  \& \bibinfo{author}{Verbiest, G.~J.}
\newblock \bibinfo{title}{Phonon scattering at kinks in suspended graphene}.
\newblock \emph{\bibinfo{journal}{Phys. Rev. B}}
  \textbf{\bibinfo{volume}{101}}, \bibinfo{pages}{115411}
  (\bibinfo{year}{2020}).

\bibitem{nihira2003temperature}
\bibinfo{author}{Nihira, T.} \& \bibinfo{author}{Iwata, T.}
\newblock \bibinfo{title}{Temperature dependence of lattice vibrations and
  analysis of the specific heat of graphite}.
\newblock \emph{\bibinfo{journal}{Phys. Rev. B}} \textbf{\bibinfo{volume}{68}},
  \bibinfo{pages}{134305} (\bibinfo{year}{2003}).

\bibitem{singh2011mechanism}
\bibinfo{author}{Singh, D.}, \bibinfo{author}{Murthy, J.~Y.} \&
  \bibinfo{author}{Fisher, T.~S.}
\newblock \bibinfo{title}{Mechanism of thermal conductivity reduction in
  few-layer graphene}.
\newblock \emph{\bibinfo{journal}{J. Appl. Phys.}}
  \textbf{\bibinfo{volume}{110}}, \bibinfo{pages}{044317}
  (\bibinfo{year}{2011}).

\bibitem{han2020ultrasoft}
\bibinfo{author}{Han, E.} \emph{et~al.}
\newblock \bibinfo{title}{Ultrasoft slip-mediated bending in few-layer
  graphene}.
\newblock \emph{\bibinfo{journal}{Nat. Mater.}} \textbf{\bibinfo{volume}{19}},
  \bibinfo{pages}{305--309} (\bibinfo{year}{2020}).

\bibitem{wei2013bending}
\bibinfo{author}{Wei, Y.}, \bibinfo{author}{Wang, B.}, \bibinfo{author}{Wu,
  J.}, \bibinfo{author}{Yang, R.} \& \bibinfo{author}{Dunn, M.~L.}
\newblock \bibinfo{title}{Bending rigidity and gaussian bending stiffness of
  single-layered graphene}.
\newblock \emph{\bibinfo{journal}{Nano lett.}} \textbf{\bibinfo{volume}{13}},
  \bibinfo{pages}{26--30} (\bibinfo{year}{2013}).

\bibitem{mercado2021impact}
\bibinfo{author}{Mercado, E.}, \bibinfo{author}{Anaya, J.} \&
  \bibinfo{author}{Kuball, M.}
\newblock \bibinfo{title}{Impact of polymer residue level on the in-plane
  thermal conductivity of suspended large-area graphene sheets}.
\newblock \emph{\bibinfo{journal}{ACS Appl. Mater. Interfaces}}
  \textbf{\bibinfo{volume}{13}}, \bibinfo{pages}{17910--17919}
  (\bibinfo{year}{2021}).

\bibitem{pettes2011influence}
\bibinfo{author}{Pettes, M.~T.}, \bibinfo{author}{Jo, I.},
  \bibinfo{author}{Yao, Z.} \& \bibinfo{author}{Shi, L.}
\newblock \bibinfo{title}{Influence of polymeric residue on the thermal
  conductivity of suspended bilayer graphene}.
\newblock \emph{\bibinfo{journal}{Nano lett.}} \textbf{\bibinfo{volume}{11}},
  \bibinfo{pages}{1195--1200} (\bibinfo{year}{2011}).

\bibitem{lindahl2012determination}
\bibinfo{author}{Lindahl, N.} \emph{et~al.}
\newblock \bibinfo{title}{Determination of the bending rigidity of graphene via
  electrostatic actuation of buckled membranes}.
\newblock \emph{\bibinfo{journal}{Nano lett.}} \textbf{\bibinfo{volume}{12}},
  \bibinfo{pages}{3526--3531} (\bibinfo{year}{2012}).

\bibitem{wang2019bending}
\bibinfo{author}{Wang, G.} \emph{et~al.}
\newblock \bibinfo{title}{Bending of multilayer van der waals materials}.
\newblock \emph{\bibinfo{journal}{Phys. Rev. Lett.}}
  \textbf{\bibinfo{volume}{123}}, \bibinfo{pages}{116101}
  (\bibinfo{year}{2019}).

\bibitem{dolleman2017amplitude}
\bibinfo{author}{Dolleman, R.~J.}, \bibinfo{author}{Davidovikj, D.},
  \bibinfo{author}{van~der Zant, H. S.~J.} \& \bibinfo{author}{Steeneken,
  P.~G.}
\newblock \bibinfo{title}{Amplitude calibration of 2\uppercase{D} mechanical
  resonators by nonlinear optical transduction}.
\newblock \emph{\bibinfo{journal}{Appl. Phys. Lett.}}
  \textbf{\bibinfo{volume}{111}}, \bibinfo{pages}{253104}
  (\bibinfo{year}{2017}).

\bibitem{maldovan2013sound}
\bibinfo{author}{Maldovan, M.}
\newblock \bibinfo{title}{Sound and heat revolutions in phononics}.
\newblock \emph{\bibinfo{journal}{Nature}} \textbf{\bibinfo{volume}{503}},
  \bibinfo{pages}{209--217} (\bibinfo{year}{2013}).

\bibitem{balandin2020phononics}
\bibinfo{author}{Balandin, A.~A.}
\newblock \bibinfo{title}{Phononics of graphene and related materials}.
\newblock \emph{\bibinfo{journal}{ACS nano}} \textbf{\bibinfo{volume}{14}},
  \bibinfo{pages}{5170--5178} (\bibinfo{year}{2020}).

\bibitem{liu20192d}
\bibinfo{author}{Liu, X.} \& \bibinfo{author}{Hersam, M.~C.}
\newblock \bibinfo{title}{2\uppercase{D} materials for quantum information
  science}.
\newblock \emph{\bibinfo{journal}{Nat. Rev. Mater.}}
  \textbf{\bibinfo{volume}{4}}, \bibinfo{pages}{669--684}
  (\bibinfo{year}{2019}).

\bibitem{chou2021acoustic}
\bibinfo{author}{Chou, Y.-Z.}, \bibinfo{author}{Wu, F.}, \bibinfo{author}{Sau,
  J.~D.} \& \bibinfo{author}{Sarma, S.~D.}
\newblock \bibinfo{title}{Acoustic-phonon-mediated superconductivity in
  rhombohedral trilayer graphene}.
\newblock \emph{\bibinfo{journal}{Phys. Rev. Lett.}}
  \textbf{\bibinfo{volume}{127}}, \bibinfo{pages}{187001}
  (\bibinfo{year}{2021}).

\bibitem{cao2018unconventional}
\bibinfo{author}{Cao, Y.} \emph{et~al.}
\newblock \bibinfo{title}{Unconventional superconductivity in magic-angle
  graphene superlattices}.
\newblock \emph{\bibinfo{journal}{Nature}} \textbf{\bibinfo{volume}{556}},
  \bibinfo{pages}{43--50} (\bibinfo{year}{2018}).

\bibitem{shi2020electronic}
\bibinfo{author}{Shi, Y.} \emph{et~al.}
\newblock \bibinfo{title}{Electronic phase separation in multilayer
  rhombohedral graphite}.
\newblock \emph{\bibinfo{journal}{Nature}} \textbf{\bibinfo{volume}{584}},
  \bibinfo{pages}{210--214} (\bibinfo{year}{2020}).

\bibitem{cai2010thermal}
\bibinfo{author}{Cai, W.} \emph{et~al.}
\newblock \bibinfo{title}{Thermal transport in suspended and supported
  monolayer graphene grown by chemical vapor deposition}.
\newblock \emph{\bibinfo{journal}{Nano lett.}} \textbf{\bibinfo{volume}{10}},
  \bibinfo{pages}{1645--1651} (\bibinfo{year}{2010}).

\bibitem{li2017measurement}
\bibinfo{author}{Li, Q.-Y.} \emph{et~al.}
\newblock \bibinfo{title}{Measurement of specific heat and thermal conductivity
  of supported and suspended graphene by a comprehensive raman optothermal
  method}.
\newblock \emph{\bibinfo{journal}{Nanoscale}} \textbf{\bibinfo{volume}{9}},
  \bibinfo{pages}{10784--10793} (\bibinfo{year}{2017}).

\bibitem{atalaya2008continuum}
\bibinfo{author}{Atalaya, J.}, \bibinfo{author}{Isacsson, A.} \&
  \bibinfo{author}{Kinaret, J.~M.}
\newblock \bibinfo{title}{Continuum elastic modeling of graphene resonators}.
\newblock \emph{\bibinfo{journal}{Nano lett.}} \textbf{\bibinfo{volume}{8}},
  \bibinfo{pages}{4196--4200} (\bibinfo{year}{2008}).

\bibitem{lopez2015increasing}
\bibinfo{author}{Lopez-Polin, G.} \emph{et~al.}
\newblock \bibinfo{title}{Increasing the elastic modulus of graphene by
  controlled defect creation}.
\newblock \emph{\bibinfo{journal}{Nat. Phys.}} \textbf{\bibinfo{volume}{11}},
  \bibinfo{pages}{26--31} (\bibinfo{year}{2015}).

\bibitem{lindsay2011flexural}
\bibinfo{author}{Lindsay, L.}, \bibinfo{author}{Broido, D.} \&
  \bibinfo{author}{Mingo, N.}
\newblock \bibinfo{title}{Flexural phonons and thermal transport in multilayer
  graphene and graphite}.
\newblock \emph{\bibinfo{journal}{Phys. Rev. B}} \textbf{\bibinfo{volume}{83}},
  \bibinfo{pages}{235428} (\bibinfo{year}{2011}).

\bibitem{nika2017phonons}
\bibinfo{author}{Nika, D.~L.} \& \bibinfo{author}{Balandin, A.~A.}
\newblock \bibinfo{title}{Phonons and thermal transport in graphene and
  graphene-based materials}.
\newblock \emph{\bibinfo{journal}{Rep. Prog. Phys.}}
  \textbf{\bibinfo{volume}{80}}, \bibinfo{pages}{036502}
  (\bibinfo{year}{2017}).

\bibitem{cong2019probing}
\bibinfo{author}{Cong, X.} \emph{et~al.}
\newblock \bibinfo{title}{Probing the acoustic phonon dispersion and sound
  velocity of graphene by raman spectroscopy}.
\newblock \emph{\bibinfo{journal}{Carbon}} \textbf{\bibinfo{volume}{149}},
  \bibinfo{pages}{19--24} (\bibinfo{year}{2019}).

\bibitem{gunst2017flexural}
\bibinfo{author}{Gunst, T.}, \bibinfo{author}{Kaasbjerg, K.} \&
  \bibinfo{author}{Brandbyge, M.}
\newblock \bibinfo{title}{Flexural-phonon scattering induced by electrostatic
  gating in graphene}.
\newblock \emph{\bibinfo{journal}{Physical Review Letters}}
  \textbf{\bibinfo{volume}{118}}, \bibinfo{pages}{046601}
  (\bibinfo{year}{2017}).

\bibitem{hunterinterface}
\bibinfo{author}{Hunter, N.} \emph{et~al.}
\newblock \bibinfo{title}{Interface thermal resistance between monolayer wse2
  and sio2: Raman probing with consideration of optical--acoustic phonon
  nonequilibrium}.
\newblock \emph{\bibinfo{journal}{Adv. Mater. Interfaces}}
  \bibinfo{pages}{2102059}.

\bibitem{zobeiri2021direct}
\bibinfo{author}{Zobeiri, H.}, \bibinfo{author}{Hunter, N.},
  \bibinfo{author}{Wang, R.}, \bibinfo{author}{Wang, T.} \&
  \bibinfo{author}{Wang, X.}
\newblock \bibinfo{title}{Direct characterization of thermal nonequilibrium
  between optical and acoustic phonons in graphene paper under photon
  excitation}.
\newblock \emph{\bibinfo{journal}{Adv. Sci.}} \textbf{\bibinfo{volume}{8}},
  \bibinfo{pages}{2004712} (\bibinfo{year}{2021}).

\bibitem{yang2020effects}
\bibinfo{author}{Yang, N.}, \bibinfo{author}{Li, C.} \& \bibinfo{author}{Tang,
  Y.}
\newblock \bibinfo{title}{Effects of chirality and stacking on the thermal
  expansion effects of graphene}.
\newblock \emph{\bibinfo{journal}{Mater. Res. Express}}
  \textbf{\bibinfo{volume}{7}}, \bibinfo{pages}{115001} (\bibinfo{year}{2020}).

\end{thebibliography}

\begin{methods}
\subsection{Sample fabrication.}
We pattern circular cavities with a depth of \SI{240}{\nano\meter} using reactive ion etching into a Si/SiO$_2$ chip. We pattern Ti/Au electrodes (\SI{5}{}/\SI{60}{\nano\meter}) on the chip firstly along with the e-beam markers, which are then used to define the cavities. The cavity depth is less than the SiO$_2$ thickness of \SI{285}{\nano\meter} to prevent an electrical short-circuit between the Si electrode and the suspended drums. We subsequently transfer large-scale 
CVD graphene over the cavities. This double-layer graphene is fabricated by stamping one monolayer CVD graphene on top of another one, where an extra layer of polymethyl methacrylate (PMMA) is attached on each graphene layer. Finally, we remove the PMMA by annealing the devices in a furnace at a pressure of 500 Torr with a constant flow of 0.5 SLPM of an inert dry gas (Ar or N$_2$) at a temperature of \SI{300}{} $^{\circ}\text{C}$. After \SI{30}{\min} annealing, most of the PMMA has been evaporated.
\subsection{Optothermal drive.}
To measure ${\omega_0}$, $\tau$, $C_{\text{fast}}$, and $C_{\text{slow}}$ in the graphene drums, we use an interferometer to actuate and detect its motion. An intensity modulated blue laser ($\lambda=$ \SI{405}{\nano\meter}) irradiates the suspended drum resulting in a periodic heat flux\cite{dolleman2017amplitude}. The heat flux results in a motion of the drum due to the thermal expansion force. A red laser ($\lambda=$ \SI{633}{\nano\meter}) is used to detect the out-of-plane motion of the graphene drum. Also, this technique allows to measure the thermal characterizations of suspended devices without the need to calibrate the spotsize and intensity of laser as in the Raman based methods\cite{cai2010thermal,li2017measurement}. All measurements are performed at room temperature inside a vacuum chamber at 10$^{-6}$ mbar. A vector network analyzer (VNA) modulates the intensity of a blue laser at frequency $\omega$ to optothermally actuate a resonator while it analyzes the resulting intensity modulation of the red laser caused by the mechanical response of the same resonator. The red and blue laser powers used are \SI{1.2}{} and \SI{0.13}{\milli\watt} respectively, where it was verified that the resonators vibrate in the linear regime and the temperature increase due to self-heating was negligible\cite{dolleman2017optomechanics}.

\subsection{Scattering model of acoustic phonons.}
We develop this model at the sidewall between the suspended (domain 1) and supported (domain 2) graphene to evaluate thermal time constant $\tau$ (see diagram in Fig. S1a). For incoming flexural phonons (defined as mode $1z$), the transmission probability ${w}_{1z\to{2j}}(j=l,t,z)$ can be calculated as a function of the incident angle $\theta_{1z}$ (see Fig. S1b for details). Here, $c_l=$\SI{21.6}{\kilo\meter/\second} and $c_t=$\SI{16}{\kilo\meter/\second} are extracted from the linear dispersion of in-plane phonons in graphene, using\cite{atalaya2008continuum} Lame parameters $\lambda=$ \SI{48}{J/\meter}$^2$ $\mu=$ \SI{144}{J/\meter}$^2$. The total transmission coefficient $\overline{w}_{1z\to{2j}}$ is then obtained by integrating ${w}_{1z\to{2j}}$ over all incidence angles from $-{\pi}/{2}$ to ${\pi}/{2}$. We also assume kink angle $\beta=90^{\circ}$ at the edge of the membrane. Wrinkles and ripples in membrane surface can be regarded as kinks with a tiny $\beta$, of which contributions to the heat transport can be negligible since $\tau$ reduces dramatically as $\beta$ decreases\cite{dolleman2020phonon}.          
\end{methods}



\begin{addendum}
 \item P.G.S. and G.J.V. acknowledges support by the Dutch 4TU federation for the Plantenna project. H.L. acknowledges the financial support from China Scholarship Council.  M.L., M.\v{S}., H.S.J.v.d.Z. and P.G.S. acknowledge funding from the European Union’s Horizon 2020 research and innovation program under grant agreement number 881603. We thank Applied Nanolayers (ANL) for transferring CVD graphene on the chips.  

 \item[Competing Interests] The authors declare that they have no
competing financial interests.
 \item[Correspondence] Correspondence and requests for materials
should be addressed to H. Liu (email: H.Liu-7@tudelft.nl) and G. J. Verbiest (email: G.J.Verbiest@tudelft.nl). 
\end{addendum}


\begin{table}
\centering
  \caption{\textbf{Characteristics of devices D1$-$D4}. Radius $r$ , effective gap $g_0$, pretension $n_0$, effective mass $m_{\text{eff}}$ , normalized areal mass $\eta$, the second derivative of the capacitance with respect to the electrostatic deflection $\frac{\partial^2C_{\text{g}}}{\partial z_{\text{g}}^2}$, bending rigidity $\kappa$ as well as the relative power absorption $\gamma$ in the ratio of thermal expansion coefficients.}
  
  \begin{tabular}{lllllllll}
    \hline
    device & $r$ & $g_0$ & $n_0$ & $m_{\text{eff}}$ & $\eta$ & $\frac{\partial^2C_{\text{g}}}{\partial z_{\text{g}}^2}$ & $\kappa$ & $\gamma$ \\
    & (\SI{}{\micro\meter}) & (\SI{}{\nano\meter}) & (\SI{}{\newton/\meter}) & ($\times$10$^{-16}$\SI{}{\kilo\gram}) & & (\SI{}{\milli\farad \meter}$^{-2}$) & (\SI{}{eV}) & \\
    \hline
    D1 & 2.5 & 212 & 0.24 & 0.50 & 19.40 & 6.56 & 3.8 & 8.18 \\
    D2 & 2.5 & 237 & 0.34 & 0.96 & 37.10  & 9.12 & 9.4 & 6.90  \\
    D3 & 5 & 227 & 0.21 & 1.06 & 10.36 & 34.75 & 2.2 & 3.26  \\
    D4 & 5 & 237 & 0.18 & 0.39 & 3.80  & 30.35 & 0.6 & 1.29  \\
    \hline
  \end{tabular}
\end{table}

\newpage

\section*{\large Supplementary Information:\\Tension tuning of sound and heat transport in graphene}


\author{H. Liu$^{1}$, M. Lee$^{2}$, M. \v{S}i\v{s}kins$^{1,2}$, H. S. J. van der Zant$^{2}$, P. G. Steeneken$^{1,2}$ and G. J. Verbiest$^1$}

\begin{affiliations}
 \item Department of Precision and Microsystems Engineering, Delft University of Technology
 \item Kavli Institute of Nanoscience, Delft University of Technology
\end{affiliations}

\section{Sample characterization}

We use indentation measurements with an atomic force microscope (AFM) to determine the 2D Young's modulus $Et$ of the suspended membranes. This indentation measurement is modelled as a clamped circular membrane with central point loading. The relationship between the applied force $F$ with the AFM cantilever and the resulting deformation $\delta$ is given by $F=n_0\pi\delta+Etq^3{\delta^3}/{r^2}$, where $q=1/(1.05-0.15\nu-0.16\nu^2)$ is a geometrical factor with a Poisson's ratio $\nu=0.16$ (ref.\cite{lopez2015increasing}) and $n_0$ is the pretension in the membrane. We extract $Et$ of our graphene devices through fitting the measured curves of $F$ vs $\delta$. A statistical analysis over 21 different drums in yielded a mean value of $Et=$\SI{175.39}{\newton/\meter} (ref.\cite{vsivskins2020sensitive}), which is therefore used in equation (3) of the main text to further estimate the induced tension.

\section{Readout of the thermodynamic properties}

As described in Methods section of the main text, an optomechanical drive allows us to actuate the graphene resonators and measure their thermodynamic properties. For the detection of the motion of the graphene membranes, we use a red laser ($\lambda$=\SI{633}{\nano\meter}) with a power of \SI{1.2}{\milli\watt}, whereas we use an intensity modulated blue laser ($\lambda$=\SI{405}{\nano\meter}) with a power $P_{ac}$ of \SI{0.13}{\milli\watt}. We sweep the frequency $\omega/2\pi$ of the intensity modulation from \SI{100}{\kilo\hertz} to \SI{100}{\mega\hertz}. In addition, to correct for intrinsic phase shifts from the interferometric setup, we calibrate the measured signals on the VNA by pointing the blue laser directly onto the photodiode (see more details in ref.\cite{dolleman2017optomechanics}). This correction allows us to obtain the real and imaginary part of the membrane motion $z_\omega$ as shown in Fig. 1e$-$1g.

The thermal peak observed in the motion of the measured graphene drums are only present in case of an optothermal drive. For other actuation methods, such as an electrostatic and piezoelectric drive, this peak is absent. Also, we reported that the used laser powers in this work have no effect on $\tau$\cite{dolleman2017optomechanics,dolleman2018transient}.

\section{Scattering model for acoustic phonons}

Acoustic phonons scatter at the kink present in the boundary of the drum with the substrate. The scattering rates depend on the tension in the membrane, speed of the phonons, and the incidence angle. In our experiment, we tune the tension and speed of the phonons and thereby also the measured $\tau$. Fig. S1a illustrates the situation for an incoming flexural phonon at the kink. Domain 1 and domain 2 represent the suspended membrane and the sidewall, respectively. The transmission probabilities for the different phonon modes ($2j$, $j=l,t,z$) at a given incident angle $\theta$ is given by
\renewcommand{\theequation}{S\arabic{equation}}
\begin{equation}
\label{eq S1}
{w}_{1z\to{2j}}=\frac{{c_{j}|u_{2j}|^2\text{Re}(\cos\theta_{2j})}}{{c_z |u_{1z}|^2 \cos\theta_{1z}}}, j=l,t,z
\end{equation}
where $|u_{1z}|$ and $|u_{2j}|$ are wave amplitudes of modes $1z$ and $2j$, respectively. The transmission angle $\theta_{2j}$ is determined by Snell's law as $\sin\theta_{2j}=c_{2j}/c_{1z}\sin\theta_{1z}$. We assume a kink angle $\beta=90^{\circ}$, in which case $|u_{2j}|$ depends both on tension $n$ and on the speed of flexural phonon $c_z$ (see derivations in ref.\cite{dolleman2020phonon}).   

\renewcommand{\thefigure}{S1}
\begin{figure}[htp]
	\centering
	\includegraphics[width=0.9\linewidth,angle=0]{"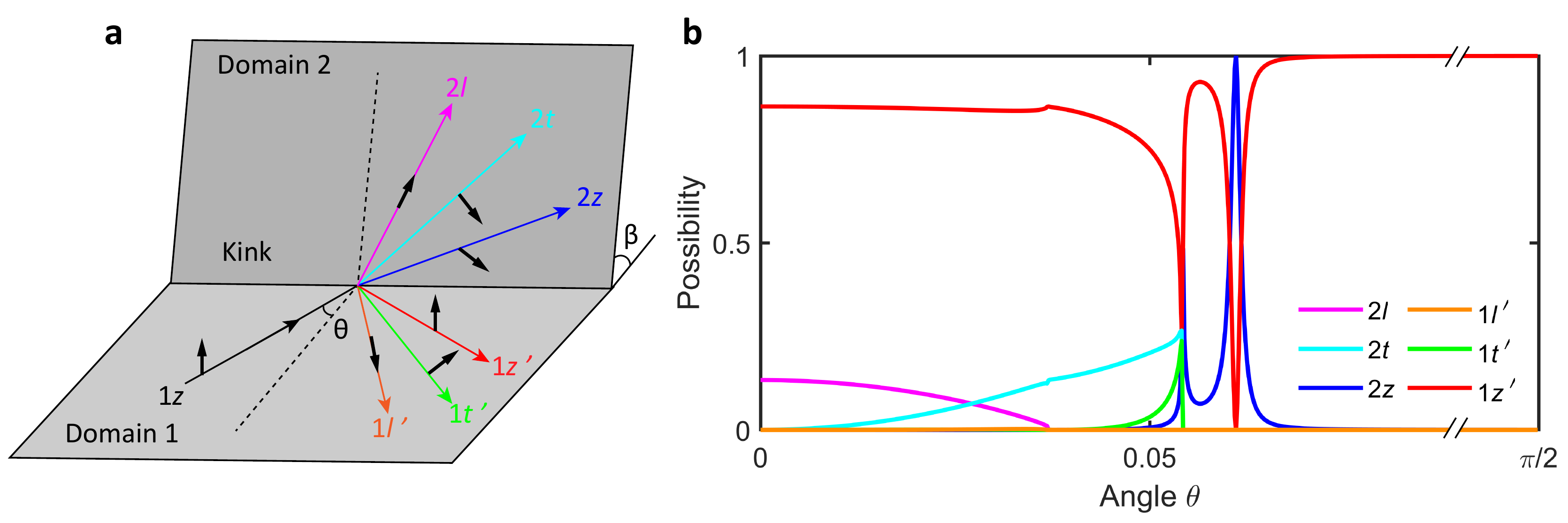"}
	\caption{\textbf{Phonons scattering at kink in membrane}. \textbf{a}, Diagram of the scattering model for an incoming flexural phonon (mode $1z$) on the kink. Transmitted and reflected modes are denoted by ($2l$, $2t$, and $2z$) and ($1l^{\prime}$, $1t^{\prime}$, and $1z^{\prime}$), respectively. \textbf{b}, Transmission and reflection possibilities as the function of the angle $\theta$ of the incoming flexural phonon computed by the full set of equations given in ref.\cite{dolleman2020phonon}. In this example, we assumed $n=$ \SI{0.3}{\newton/\meter} and $c_z=$ \SI{575}{\meter/\second}.}
	\label{fig:capture-S1}
\end{figure}

Using equation (S1), we plot all transmission and reflection possibilities versus $\theta$ (Fig. S1b) for a tension of $n=$ \SI{0.3}{\newton/\meter} and $c_z=$ \SI{575}{\meter/\second}. The graph shows that transmission of the incoming flexural phonon to the sidewall only occurs when $\theta<4.29^{\circ}$. Interestingly, mode $2z$ (blue line) exhibits two peaks at around $3.12^{\circ}$ and $3.50^{\circ}$, which are attributed to a resonant transmission of waves. On the other hand, since ${w}_{1z\to{1z^{\prime}}}\simeq1$ when $\theta>4.29^{\circ}$, the incident flexural phonons are almost always reflected, which results in a large thermal boundary resistance. We then calculate transmission coefficients $\overline{{w}}_{1z\to{2j}}$ ($j=l,t,z$) through integrating ${{w}}_{1z\to{2j}}$ over the range of $\theta$ from $-\pi/2$ to $\pi/2$. Finally, we compute $\tau$ as the functions of $c_z$ and $n$ by substituting $\overline{{w}}_{1z\to{2j}}$ into equation (3) in the main text. 

\renewcommand{\thefigure}{S2}
\begin{figure}
	\centering
	\includegraphics[width=1\linewidth,angle=0]{"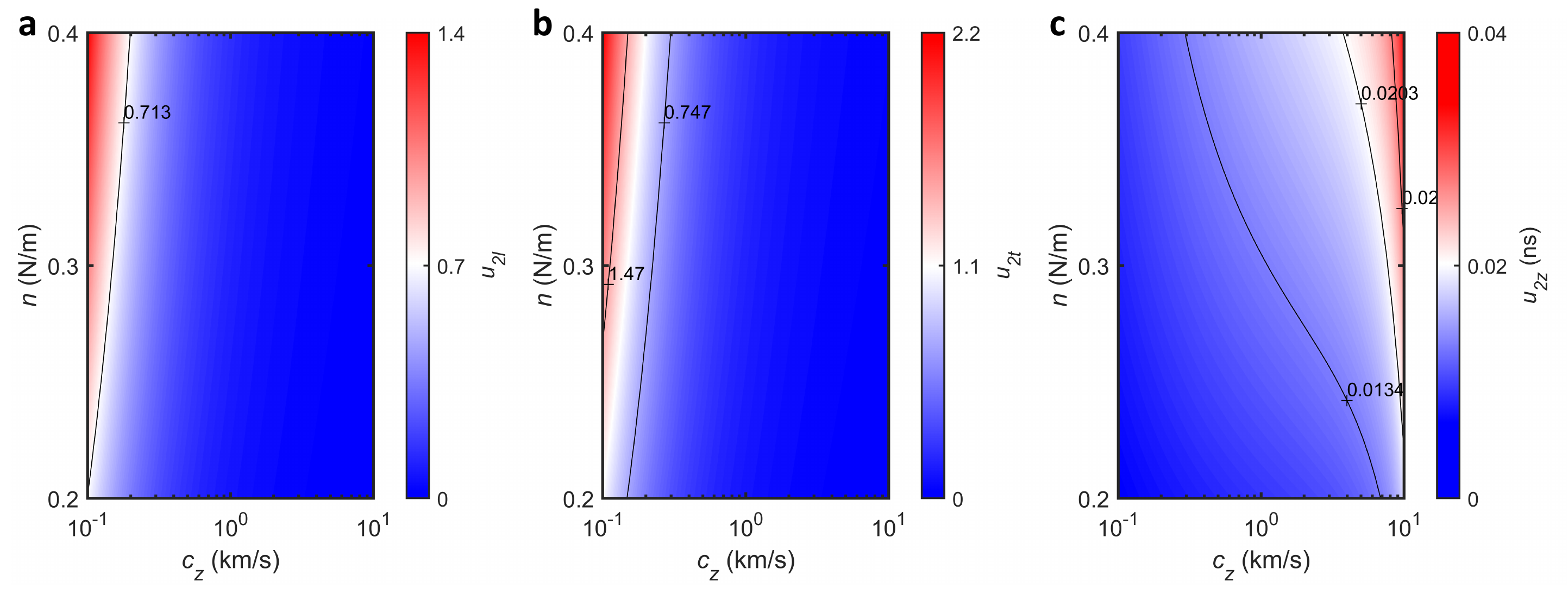"}
	\caption{\textbf{Transmitted wave amplitudes of three types of acoustic phonons versus the sound speed of flexural phonons $c_z$ and tension $n$.} \textbf{a$-$c}, Results of $u_{2l}(n,c_z)$, $u_{2t}(n,c_z)$ and $u_{2z}(n,c_z)$, respectively.}
	\label{fig:capture-S2}
\end{figure}

Using the expression of wave amplitudes of the transmitted modes 2$j$ (ref.\cite{dolleman2020phonon}), we further discuss the dependence of $n$ and $c_z$ on $u_{2l}$, $u_{2l}$ and $u_{2l}$, respectively (see Fig. S2). E.g., as tension $n$ increases, the impedance matching between the flexural phonons on suspended membrane and the in-plane phonons on supported membrane is improved, due to a weakening of deflection-induced dilatation and shear, results in the growing amplitudes of $u_{2z}$ mode and further play the role on $\tau$. The increase of $c_z$ will reduce $u_{2l}$ and $u_{2t}$, and only contribute positively to the flexural mode $u_{2z}$. In general, $u_{2l}$ and $u_{2t}$ are two-order-larger than $u_{2z}$, showing the intense reflection of flexural phonons at the kink.

\renewcommand{\thefigure}{S3}
\begin{figure}
	\centering
	\includegraphics[width=1\linewidth,angle=0]{"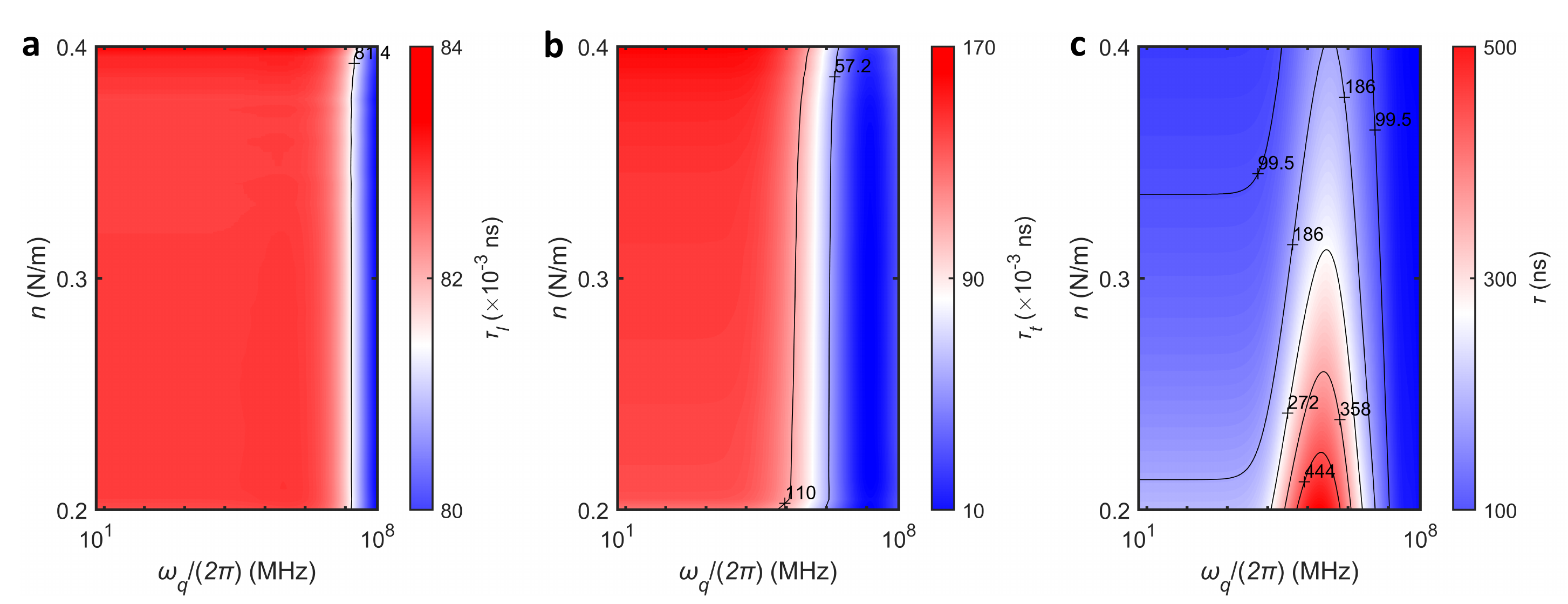"}
	\caption{\textbf{Thermal time constants of three types of acoustic phonons versus the phonons frequency $\omega_q/2\pi$ and tension $n$.} \textbf{a$-$c}, Results of $\tau_l(n,\omega_q)$, $\tau_t(n,\omega_q)$ and $\tau(n,\omega_q)$, respectively. In general, the magnitude of $\tau(n,\omega_q)$ is three order larger than that of $\tau_l(n,\omega_q)$ and $\tau_t(n,\omega_q)$, indicating the dominant contribution of flexural phonon on the thermal characterization in graphene. The calculated result in \textbf{c} is then weighted by the $\omega_q$-dependent specific heat and obtain the total $\tau(n)$ in theory.}
	\label{fig:capture-S3}
\end{figure}


Note that equation (S1) and equation (2) in the main text are not only applicable for calculating $\tau$, but also for calculating the thermal time constants $\tau_l$ (LA) and $\tau_t$ (TA) of the in-plane phonons. Also realize that the speed of sound $c_z$ is a function of the phonon frequency $\omega_q$ (see SI section 4) through the dispersion relation of the material. To illustrate these dependencies, we plot $\tau_l(n,\omega_q)$, $\tau_t(n,\omega_q)$ and $\tau(n,\omega_q)$ as a function of both $\omega_q$ and $n$ (see Fig. S3). In this example, the parameters of device D4 are used, with a drum radius of $r$ is \SI{5}{\micro\meter} and a bending rigidity $\kappa$ of \SI{0.6}{eV}. As Figs. S3a and S3b show, both $\tau_l$ and $\tau_t$ decrease as $\omega_q$ increases due to Snell’s law, but are nearly independent of $n$. On the contrary, $\tau$ exhibits a strong dependence on $n$ (Fig. S3c). Generally speaking, $\tau(n,\omega_q)$ is roughly three orders of magnitude higher than $\tau_l(n,\omega_q)$ and $\tau_t(n,\omega_q)$, which agrees well with flexural phonon dominated the heat transport in graphene\cite{lindsay2010flexural,lindsay2011flexural}. The measured $\tau$ in our graphene drum devices is therefore equal to weighed average over the contributions of all flexural phonons with frequency $\omega_q$.

\section{Debye model of acoustic phonons}

Acoustic phonons in graphene exhibit a dispersion relations \cite{nika2017phonons}. In-plane (LA and TA) phonons have a linear dispersion relation $\omega_q=c_jq$ ($j=l, t$), in which $q$ is the wavenumber (see Fig. S4a, left panel). In line with theoretical and experimental work, we use $c_l=$\SI{21.6}{\kilo\meter/\second} and $c_t=$\SI{16}{\kilo\meter/\second} in our calculations\cite{nihira2003temperature,cong2019probing}. On the other hand, flexural acoustic phonons have a nonlinear dispersion relation\cite{gunst2017flexural} expressed as $\omega_q=\sqrt{(\kappa q^4+nq^2)/(\eta\rho_g)}$, where $\kappa$ is the bending rigidity of the membrane, and $\eta$ is the normalized areal mass of the membrane extracted from the effective mass density $m_\text{eff}$ (obtained from the fits in Fig. 2a). The speed of flexural phonons is thus expressed as $c_z=\frac{\partial \omega_q}{\partial q}$. Furthermore, we can obtain the specific heat spectral density $C_{v,\omega}^{z}$ of flexural phonons in graphene\cite{nihira2003temperature}:
\renewcommand{\theequation}{S\arabic{equation}}
\begin{equation}
\label{eq S2}
C_{v,\omega}^z = k_B \left( \frac{\hbar\omega_q}{kT} \right)^2\frac{e^{\hbar\omega_q/kT}}{\left(e^{\hbar\omega_q/kT}-1\right)^2} D\left(\omega_q\right)
\end{equation}
where $k_B$ is the Boltzmann constant, $\hbar$ is the Planck constant divided by $2\pi$, and $D(\omega_q)$ is the density of states for the given dispersion relations. Using the above equation (S2) We can also obtain the specific heat spectral density $C_{v,\omega}^j$ ($j=l, t$) of in-plane acoustic phonons (see Fig. S4a, right panel), which are independent to any parameters and remain unchanged in our work.  

For the flexural dispersion $\omega_q$, $\kappa$ dominates the high frequency (THz) regime while $n$ dominates the low frequency (MHz) regime. We define the point where the domination transmits as the cross-over frequency $\omega_{qc}$ of dispersion. Through intersecting $\omega_q=q\sqrt{n/(\eta\rho_g)}$ and $\omega_q=q^2\sqrt{\kappa/(\eta\rho_g)}$, we determine $\omega_{qc}$ for all devices D1$-$D4, located at 84.8, 52.6, 174.4 and 422.7~GHz, respectively (see Fig. S4b). 

\renewcommand{\thefigure}{S4}
\begin{figure}
	\centering
	\includegraphics[width=0.75\linewidth,angle=0]{"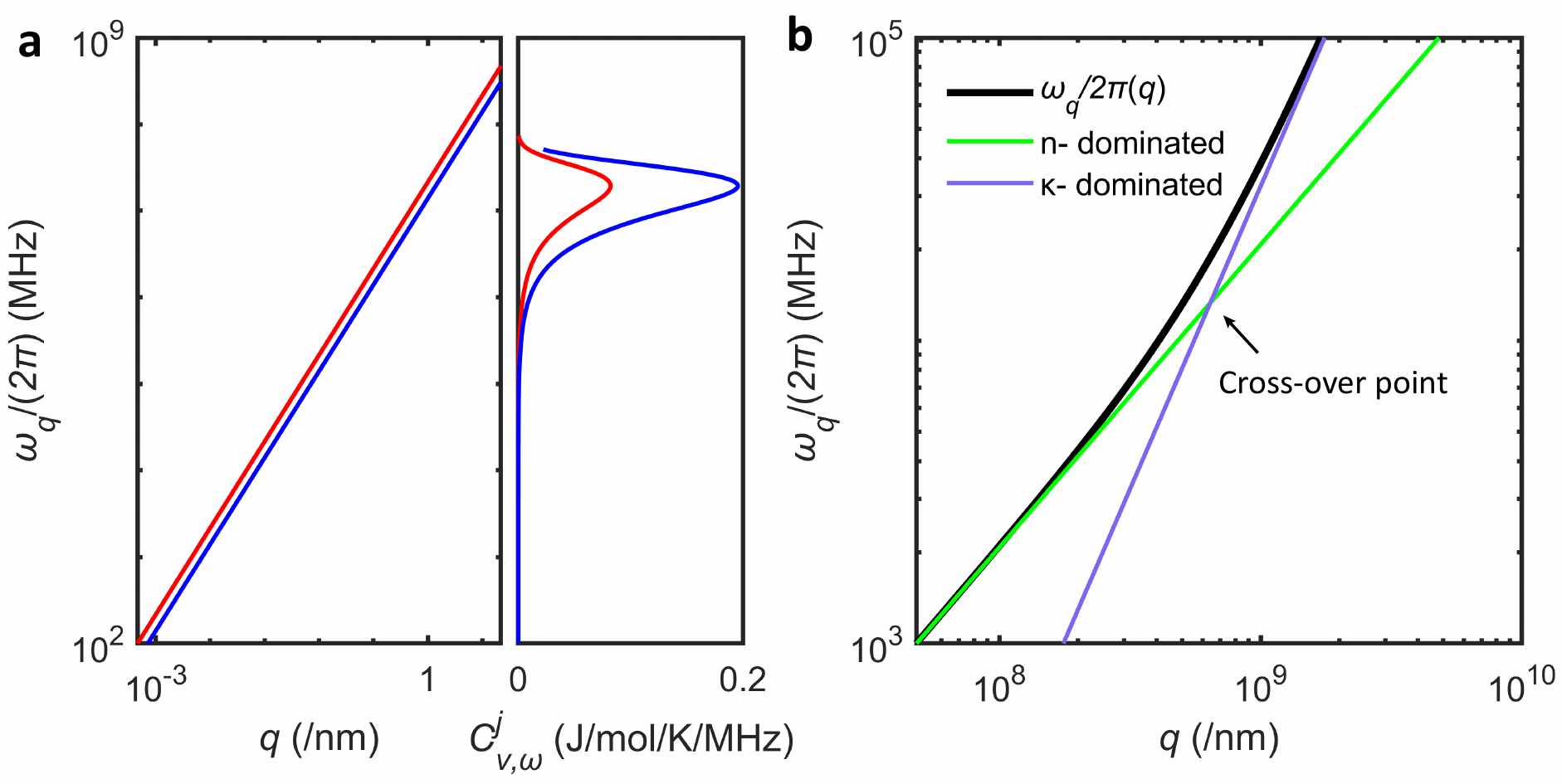"}
	\caption{\textbf{Calculations for dispersion relation of acoustic phonons.} \textbf{a}, Left panel: dispersion relation of in-plane acoustic phonons LA (red line) and TA (blue line); right panel: the corresponding specific heat spectral density $C_{v,\omega}^j$ ($j=l, t$). \textbf{b}, Dispersion relation of flexural phonons for device D1 (black line), using the extracted tension $n=$ \SI{0.25}{N/m}, bending rigidity $\kappa=$ \SI{3.8}{eV} and normalized areal mass $\eta=19.401$ from the main text. The cross-over frequency $\omega_{qc}$ corresponds to the transition of dispersion relation from $n$- dominated regime to $\kappa$- dominated regime.}
	\label{fig:capture-S4}
\end{figure}

Let us now explore the roles $\kappa$, $\eta$ and $n$ play on our Debye scattering model (see Fig. S5). Assume the initial values of these factors as $\kappa=$ \SI{0.6}{eV}, $\eta=10$ and $n=0.3$ \SI{}{N/m}, as well as the radius of membrane $r=$ \SI{5}{\micro\meter}. In terms of $\omega_q(q)$ and $c_z$, $\kappa$ and $n$ mainly play the roles on THz and MHz regime, respectively, while $\eta$ does on both regimes (see Fig. S5a-f). On the other hand, $\kappa$ and $n$ purely tune the magnitude and the $\omega_q$- distribution of $C_{v,\omega}^{z}$, separately, while they act on the inverse way for $\tau$ (see Fig. S5g-l). It should be noted that the sensitivity of $n$ on $\tau$ is extremely higher compared to that of $\kappa$ and $\eta$. Hence, the tunability of $\tau$ as observed in our measurements is mainly attributed to the induced $n$, which directly contributes to the impedance matching of the acoustic phonons at the edge of the membrane.                 

For each $\omega_{q}$, the above calculation gives us the thermal time constant and the specific heat spectral density for all types of acoustic phonons and a given tension $n$. Following the definition of the specific heat, we use a weighing relation as introduced, ${1}/{\tau}= \int_{0}^{\omega_{qd}} C_{v,\omega}^i({\omega_q})/({C_v^i \tau(\omega_q)}) \text{d}\omega_q$, where $C_v^i=\int_{0}^{\omega_{qd}} C_{v,\omega}^{i}({\omega_q}) \text{d}\omega_q$ is the total specific heat of a particular phonon type $i = {l,t,z}$, to separately obtain $\tau$, $\tau_l$ and $\tau_t$. Following the flow chart depicted in Fig. 3a of the main manuscript, $\kappa$ is the only fitting parameter to match the calculated to the measured $\tau$. 

\renewcommand{\thefigure}{S5}
\begin{figure}
	\centering
	\includegraphics[width=1\linewidth,angle=0]{"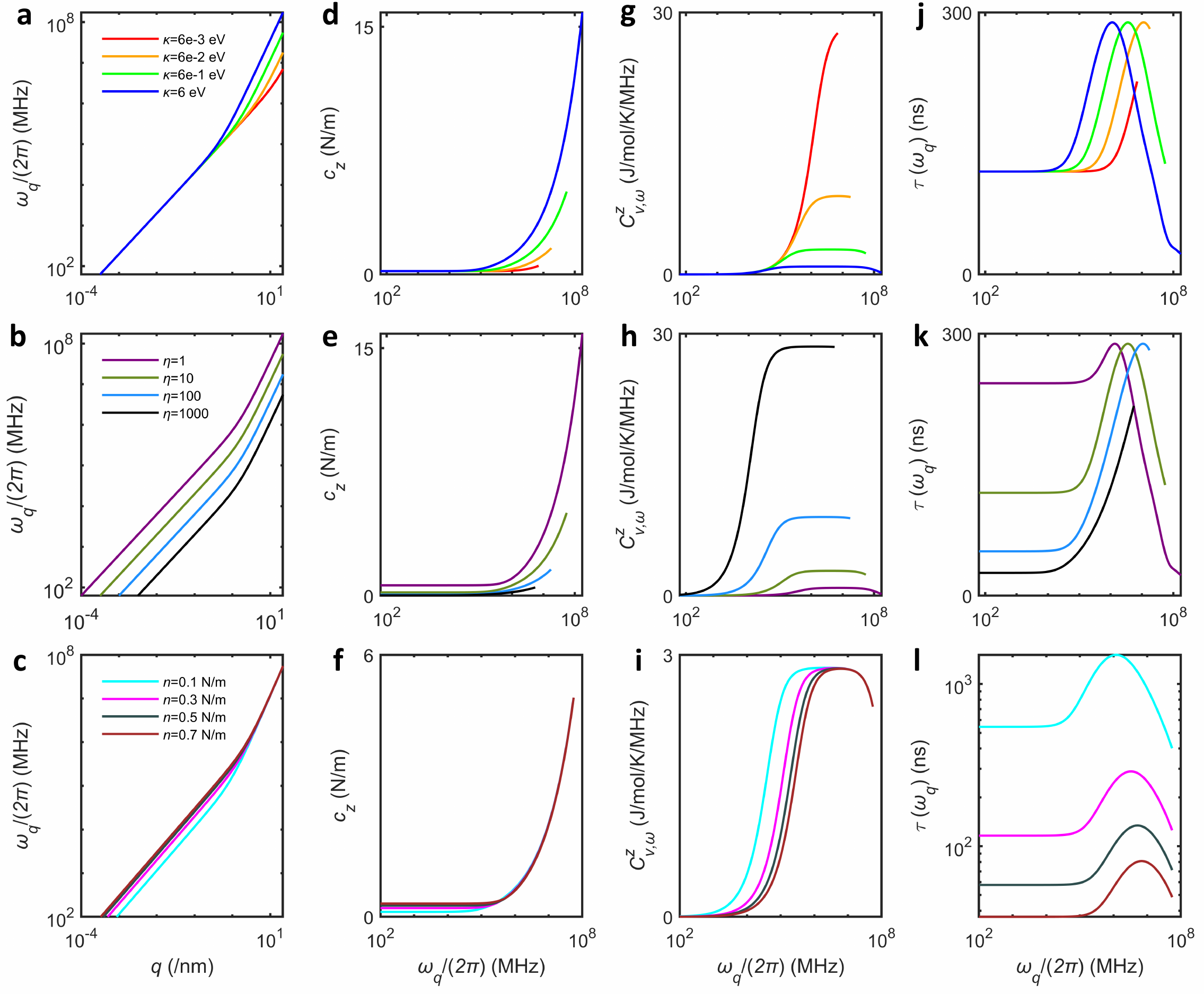"}
	\caption{\textbf{Discussion for Debye scattering model.} Four parameters of flexural acoustic phonons, including frequency $\omega_{q}$, speed $c_z$, specific heat spectral density $C_{v,\omega}^{z}$ and thermal time constant $\tau$ are discussed with respect to \textbf{a}, bending rigidity $\kappa$; \textbf{b}, normalized areal mass $\eta$; \textbf{c}, tension $n$. Initial settings are $\kappa=$ \SI{0.6}{eV}, $\eta=10$ and $n=0.3$ \SI{}{N/m}, and the radius of membrane $r=$ \SI{5}{\micro\meter}.}
	\label{fig:capture-S5}
\end{figure}

We further discuss the role of $\kappa$ played on the tension-dependent tunable $\tau$ (Fig. S6). It shows that as $\kappa$ goes up from 0.6 to \SI{19}{eV}, the magnitude of $\tau$ is improved obviously. This can be explained that the increase of $\kappa$ leads to the decrease of $c_z$ (Fig. S5d) as well as the increase of $C_{v,\omega}^{z}$ (Fig. S5d). The former will cause the decrease of transmitted amplitude in scattering model (see Fig. S2), while the latter will directly improve the weighed factors of $\tau$, both of which will result in the enhancement of $\tau$. In addition, we also observe the slope $|\partial\tau/\partial n|$ increases with $\kappa$ at the same moment.      

\renewcommand{\thefigure}{S6}
\begin{figure}
	\centering
	\includegraphics[width=0.4\linewidth,angle=0]{"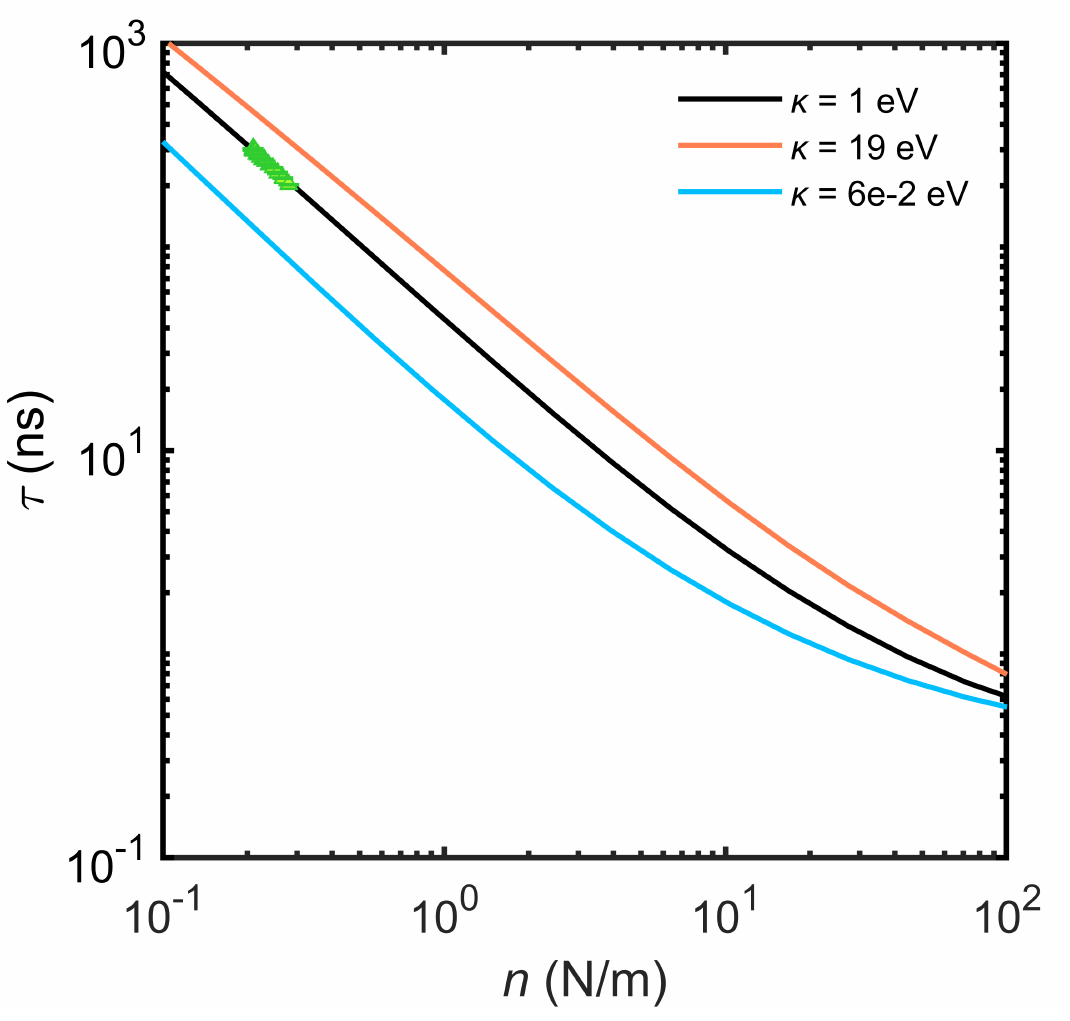"}
	\caption{\textbf{Influence of bending rigidity on tension-dependent thermal time constant.} Green points, measured thermal time constant $\tau$ as a function of tension $n$ for device D3 in the main manuscript; lines, estimated $\tau$ versus $n$ under different values of $\kappa$.}
	\label{fig:capture-S6}
\end{figure}

\section{Thermal nonequilibrium of acoustic phonons}

Characterizing thermal nonequilibrium among phonons in graphene has recently received a remarkable attention\cite{hunterinterface,wang2020distinguishing,zobeiri2021direct}. In this work, we consider the absolute value of the thermal expansion forces ratio $|C_{\text{fast}}/C_{\text{slow}}|$. According to the 2D heat equation\cite{dolleman2017optomechanics}, the thermal expansion amplitude is equal to $\alpha P_{\text{abs}}R$, where $\alpha$ and $R$ is the thermal expansion coefficient and thermal resistance of acoustic phonons, and $P_{\text{abs}}$ is the absorbed laser power by the phonons. Assume the values of $\alpha$ are equivalent for all acoustic phonons\cite{yang2020effects}, using $\tau=RC_v$, we have $|C_{\text{fast}}/C_{\text{slow}}| = \gamma \frac{(\tau_l/C_v^l+\tau_t/C_v^t)}{\tau/C_v^z}$, where $\gamma = P_{\text{abs}}^{{l,t}}/P_{\text{abs}}^{{z}}$ represents the power absorption ratio between in-plane and flexural phonons. Through the fitting of $\gamma$, we obtain a good match between the computed and measured tension-dependence of $|C_{\text{fast}}/C_{\text{slow}}|$ for all devices (see Fig. 3c in the main manuscript). We express the relative power absorption $\xi$ by flexural phonon as $\xi=1/(\gamma+1)$, which is further plotted as the function of $\eta$ (see Fig. 3d in the main manuscript) to analyze the effect of residues.

\end{document}